\documentclass[review,number,sort&compress]{elsarticle}
\usepackage{lineno}
 \linenumbers 
\usepackage{multirow}
\usepackage{amssymb}
\usepackage{graphicx}
\usepackage{amsmath}
\usepackage{amsfonts}
\usepackage{etoolbox}

\usepackage[squaren]{SIunits}
\usepackage{color}
\usepackage{xcolor}

\usepackage[utf8]{inputenc}
\usepackage[T1]{fontenc}

\newcommand{\gfrac}[2]{\displaystyle\frac{#1}{#2}}

\newcommand{\dd}{\mbox{d}}

\newcommand{\E}{\mathbb{E}}
\newcommand{\cov}{\mathrm{Cov}}
\newcommand{\RMS}{\mathrm{RMS}}

 \newcommand{\Blue}[1]{\textcolor{blue}{#1}}
 \newcommand{\Red}[1]{\textcolor{red}{#1}}
 
 \newcommand{\Green}[1]{\textcolor{green}{#1}}

 \newcommand{\Orange}[1]{\textcolor{orange}{#1}}

\journal{Nucl.\ Instrum.\ Meth.\ A}

\begin{document}

\begin{frontmatter}

\title{Charged particle tracking without magnetic field: optimal measurement of track momentum by a Bayesian analysis of the multiple measurements of deflections due to multiple scattering}

\author[add1]{Mikael Frosini}
\author[add1]{Denis Bernard\corref{cor}}
\ead{denis.bernard at in2p3.fr}
\address[add1]{LLR, Ecole Polytechnique, CNRS/IN2P3, 91128 Palaiseau, France}


\begin{abstract}
We revisit the precision of the measurement of track parameters
(position, angle) with optimal methods in the presence of detector
resolution, multiple scattering and zero magnetic field.
We then obtain an optimal estimator of the track momentum by a
Bayesian analysis of the filtering innovations of a series of Kalman
filters applied to the track.

This work could pave the way to the development of autonomous
high-performance gas time-projection chambers (TPC) or silicon wafer
$\gamma$-ray space telescopes and be a powerful guide in the
optimisation of the design of the multi-kilo-ton liquid argon TPCs
that are under development for neutrino studies.
\end{abstract}


\end{frontmatter}


\section{Introduction}
\label{sec:introduction}

\subsection{$\gamma$-ray astronomy}
\label{sub:sec:intro:gamma:astro}

A huge effort is in progress to design $\gamma$-ray telescopes able to
bridge the sensitivity gap that extends between the upper end of the
high-sensitivity energy range of past and present X-ray telescopes and
the lower end of the high-sensitivity energy range of the Fermi-LAT
telescope, that is, approximately $0.1 - 100\,\mega\electronvolt$.

On the low-energy side of the gap, tracking of the electron issued
from the first Compton scattering of an incident photon enables a
major improvement of the precision of the reconstruction of the
direction of the incident photon (\cite{Tanimori:2017ihu} and
references therein) that induces an impressive improvement of the
true-photon-background rejection and therefore of the
point-like-source sensitivity.
A serious limitation of that ETCC (electron tracking Compton camera)
scheme arises though, as the effective area undergoes a sharp drop for
photon energies above $0.5\,\mega\electronvolt$, due to the fact that
the recoil electron can exit on the side and escape energy measurement
\cite{Tanimori:2017ihu}: electron momentum measurement inside the
time projection chamber (TPC) itself is highly desirable.

On the high-energy side of the gap, novel approaches improve the
sensitivity by improving the single-photon angular resolution by using
converters having a lower-$Z$ than that of the tungsten plates of the
EGRET / Fermi-LAT series.
Using a series of silicon wafer active targets placed at a distance
of each other, at the same time the material in which the photon
converts and in which the tracks are tracked, enables an improvement
of $\approx$ a factor of three in the angular resolution at
$100\,\mega\electronvolt$ with respect to the LAT
\cite{TIGRE:2001,MEGA:2005,CAPSiTT:2010,Morselli:2014fua,Moiseev:2015lva,AMEGO,E-Astrogam:2016,Wu:2014tya}
at the cost of a lower average active target density.
Similar values of the angular resolution are achieved using a
high-spatial-resolution, homogeneous, high-density material such as an
emulsion \cite{Takahashi:2015jza}.

If the trend to lower densities is pushed to the use of
a gaseous detector, the angular resolution with respect to the LAT
can increase up to a factor of ten at $100\,\mega\electronvolt$
\cite{Bernard:2012uf} and the single-track angular resolution is so
good that the azimuthal angle of the $e^+e^-$ pair can be measured with
a good enough precision to enable the measurement of the linear
polarization fraction of the incident radiation
\cite{Bernard:2013jea,Gros:SPIE:2016,Gros:2016dmp}.
Gas detectors enable the detection of low-energy photons close to the
pair-creation threshold where most of the statistics lie for cosmic
sources (Fig. 1 of \cite{Bernard:2017xov}), something which is
critical for polarimetry.

Astrophysicists also need to measure the energy of incoming photons
and therefore the momentum of the conversion electron(s).
This can be achieved using a number of techniques.

\begin{itemize}
\item
In a calorimeter, the total energy of the photon is absorbed and measured.
\item
In a magnetic spectrometer, the trajectory of a particle with electric
charge $q$ and momentum $p$ in a magnetic field $B$ is curved with a
curvature radius $\rho = p / (q B)$: from a measurement of $\rho$, one
obtains a measurement of $p$ and in the end of the photon energy $E$.
\item
In a transition radiation detector (TRD), the energy of the radiation
emitted
in the forward direction
by a charged particle at the interface between two media that have
different refraction indices is proportional to the Lorentz factor
$\gamma$ of the particle, enabling a direct measurement.
The low number of emitted photons per track per interface has lead to
the development of multi-foil systems that suffer destructive
interference at high energies.
Appropriate configurations have showed saturation values larger
than $\gamma \approx 10^4$, which implies that a measurement can be
done up to a photon energy of $\approx 10\,\giga\electronvolt$
\cite{Wakely:2004gg}.
\end{itemize}

The low-density active targets that have been considered above can
provide a large effective area telescope only with a large volume
(${\cal O}(\meter^3)$) and therefore the mass of the additional device used for
energy measurement is a serious issue onboard a space mission.
{\bf In this document we first address the performance of the track momentum
 measurement from measurements of the angular deflections of charged
 tracks due to multiple scattering during the propagation in the
 tracker itself}.

\subsection{Large noble-liquid TPCs for neutrino physics}
\label{sub:sec:intro:TPC:neutrino}

Neutrino oscillation is a well established phenomenon and several
experiments are being prepared with the goals:
\begin{itemize}
 \item To test the occurrence of CP violation in the neutral lepton
 sector, i.e. to measure the only free complex phase $\delta$ of
 the Pontecorvo-Maki-Nakagawa-Sakata (PMNS) matrix with enough
 precision to determine its non-compatibility with zero,
 \item
 To determine unambiguously the 3 neutrino mass ordering, i.e. to solve the sign ambiguity of the square mass difference $\Delta m^2_{31}$.
 \end{itemize}
Not only the (vacuum propagation) phase term that involves $\delta$ changes sign
upon $\nu \leftrightarrow \overline\nu$ exchange, but the term that
describes the interaction with matter changes sign too as our Earth
contains much more electrons than positrons.
``In the few-GeV energy range, the asymmetry from the matter effect
increases with baseline as the neutrinos pass through more matter,
therefore an experiment with a longer baseline [is] more sensitive
to the neutrino mass hierarchy. For baselines longer than $\approx$
1200 km, the degeneracy between the asymmetries from matter and
CP-violation effects can be resolved''
\cite{Acciarri:2015uup}.
Large distances imply low fluxes, that is, huge detectors 
and, to match the $\sin{(\Delta m L / 4 E_\nu)}$ oscillation function, 
high-energy neutrinos.
So we should be prepared to measure the momentum of high-momentum muons
in huge non-magnetised detectors such as liquid argon (lAr) TPCs.

The DUNE experiment expects to be able to measure muon momenta with a relative precision of $\approx 18\%$ 
\cite{Acciarri:2016ooe}, based on a past ICARUS work \cite{Ankowski:2006ts}.
They ``anticipate that the resolution will deteriorate for
higher-energy muons because they scatter less'', though.
Given the $\dd E/\dd x$ of $0.2\,\giga\electronvolt/ \meter$ of minimum
ionising particles in lAr, a typical $6\,\giga\electronvolt/c$ muon
produces a long track:
it should be interesting to study to what extent an optimal analysis
of the thousands of measurements per track,
at their $\approx 3\,\milli\meter$ sampling pitch,
can do better.

\subsection{Track momentum measurement from multiple scattering}
\label{sub:sec:intro:momentum:measurement}

The measurement of track momentum using multiple scattering was
pioneered by Molière \cite{Moliere} and has been used since, in
particular in the context of emulsion detectors (recent accounts can
be found in \cite{Kodama:2007mw,OPERA:2011aa}).

In a practical detector consisting of $N$ detection layers, the
precision of the deflection measurement and therefore of the momentum
measurement is affected by the precision, $\sigma$, of the
measurement of the position of the track when crossing each layer: the
combined square deflection angle summed up over the whole track length
therefore includes contributions from both the scattering angle and
the detector precision.
Bernard has optimized the longitudinal ``cell'' length over which each
deflection angle is measured \cite{Bernard:2012uf} and obtains a value
of the relative momentum precision $\sigma_p / p$ that scales as
$p^{1/3}$, but the fact that the track position precision can improve
when the cell length is extended and several measurements are combined
was not taken into account in \cite{Bernard:2012uf}.
{\bf In the present document we study an optimal method of momentum
 measurement with a tracker that has a finite (non zero) precision.
}

In Section \ref{sec:tracking} we revisit optimal tracking methods in a
context where the momentum of the particle is known.
This allows us to present concepts and notations that are used later
in the paper.
We also extend the results published in the past by the use of more
powerful methods.

The optimal precision of track measurements obtained in Sec.
\ref{sec:tracking} can be obtained by performing the fit with a
Kalman filter (KF), a tool that was imported in our field by Frühwirth
\cite{Fruhwirth:1987fm}.
In section \ref{sec:kalman} we give a brief summary of Kalman filter
tracking in a Bayesian formalism.
In magnetic spectrometers, the particle momentum takes part both in
the particle state vector through the curvature of the trajectory and
in the magnitude of multiple scattering.
The precision of the magnetic measurement is most often so good that
the momentum can rightfully be considered as being perfectly known in the
expression of the multiple scattering.
In our case of a zero magnetic field, it is not the case.
A Kalman filter is the optimal estimate for linear system models
with additive white noise, such is the case for multiple
scattering (process noise) and detector precision (measurement noise),
but at the condition that the optimal Kalman gain be used in the
expression, that is, that the track momentum be known.
In section \ref{sec:momentum:measurement}, we use the Bayesian method
developed by Matisko and Havlena
\cite{Matisko:Havlena:2013}
to obtain an optimal estimator of the
process noise covariance, and therefore of the track momentum
\footnote{Attempts of estimation of track momenta based on the use of
 a Kalman filter have been performed in the past, with little
 success. The un-validated un-characterized study of
 Ref. \cite{Wu:2015wol}, for example, shows a poor relative
 resolution of $\sigma_p / p = 30 - 40 \%$ and that does not vary
 with the true particle momentum between $50\,\mega\electronvolt/c$
 and $2\,\giga\electronvolt/c$, which is a bad symptom.}.
We implement this method and characterize its performance on Monte
Carlo (MC) simulated tracks.
We check that the momentum measurement is unbiased within
uncertainties.
We obtain a heuristic analytical expression of the relative momentum
uncertainty.

Numerical examples are given for a homogeneous gas detector such as an
argon TPC and for a silicon-wafer detector:
\begin{itemize}
\item TPC gas, argon, 5 bar, $\sigma = l = 0.1\,\centi\meter$,
 $L = 30\,\centi\meter$ \cite{Bernard:2013jea};
 
\item Silicon detector $N=56$, $\Delta x = 500\,\micro\meter$-thick
 wafers spaced by $l = 1\,\centi\meter$, with a single point
 precision of $\sigma = 70\,\micro\meter$ \cite{E-Astrogam:2016}.
\end{itemize}

In this work a number of approximations are done: only the Gaussian
core of the multiple-scattering angle distribution is considered and
the non-Gaussian tails due to large-angle single scatters are neglected.
The small logarithmic correction term in the expression of the RMS
multiple scattering angle, $\theta_0$, is neglected
\begin{equation}
\theta_0 \approx \gfrac{p_0}{\beta p} 
\sqrt{\gfrac{\Delta x}{X_0}},
 \label{eq:multiple:scattering:base}
\end{equation}
where $p_0 = 13.6\,\mega\electronvolt/c$ is the
``multiple-scattering constant'', $\Delta x$ is the matter thickness through
which the particle propagates and $X_0$ is its radiation length
(Eqs. (33.14), (33.15), (33.17) of \cite{Olive:2016xmw}).
In the case of a homogeneous detector, the thickness of the scatterer
is equal to the length of the longitudinal sampling, $l = \Delta x$.
We assume relativistic particles ($\beta \approx 1$) without loss of generality.
Only the first-order term (angle deflection) of multiple scattering is
taken into account which is legitimate for the thin detectors
considered here;
the 2nd-order transverse displacement (eq.~(33.19) of \cite{Olive:2016xmw}) is neglected. 
Continuous ($\dd E / \dd x$) and discrete (BremsStrahlung radiation) 
energy losses are also neglected.
In TPCs in which the signal is sampled, most often the electronics
applies a shaping of the pulse before digitisation, that creates a
short scale longitudinal correlation between successive measurements
that we neglect too.
Also the limitations of pattern recognition, that is, in the case of
$\gamma$-ray telescopes, of the assignment of each hit to one of two
close tracks, are not addressed.

Note that in the two main parts of this work (section
\ref{sec:tracking} and sections
\ref{sec:kalman}-\ref{sec:momentum:measurement}) we have made our best
to follow the notations of Refs. \cite{Billoir:1983mz,Innes:1992ge}
and of \cite{Matisko:Havlena:2013}, respectively, and that they turn out
to differ to some extent.

\section{Tracking}
\label{sec:tracking}

An optimal tracking makes use of the full $N \times N$ covariance
matrix of the $N$ measurements, including multiple scattering
(correlation terms). This is most often impractical in modern trackers
that provide a huge number of measurements for each track.
The first successful attempt to perform a recursive determination of
the covariance matrix was achieved by Billoir \cite{Billoir:1983mz}.
He considered the paraxial propagation of a charged track along the
$x$ axis inside a magnetic field oriented along $z$: close
to the particle origin, the trajectory is a straight line in the
$(x, z)$ plane, and a parabola osculatrix to the true circle in the
$(x,y)$ plane.
As we examine here the case of a magnetic-field-free tracker,
the propagation (in the $(x, z)$ and in the $(x,y)$ planes) is
approximated by straight lines (already using Innes notations
\cite{Innes:1992ge} but assuming $B = 0$):
\begin{equation}
y = a + b \times x.
 \label{eq:droite}
\end{equation}

Astronomers obviously have a special interest in the slope $b$, that is,
in the paraxial direction of the track at the conversion vertex.

The $(a,b)$ correlation matrix is named $V$ and the information matrix,
$I \equiv V^{-1}$.
Billoir develops a recursive method in which the fit propagates along
the track, adding the information gain (measurement) and loss
(scattering) at each layer.
He obtains the information matrix at layer $n+1$ from the information
matrix at layer $n$ \cite{Billoir:1983mz,Innes:1992ge}:
\begin{equation}
I_{n+1} = D^T \left(I_n^{-1} + B \right)^{-1} D + M,
 \label{eq:Billoir}
\end{equation}
where $D$ is the drift matrix that propagates the track from layer
$n$ to layer $n+1$,
\begin{equation}
 D = 
\begin{bmatrix} 1 & l \\ 0 & 1 \end{bmatrix},
 \label{eq:Billoir:drift}
\end{equation}
$l$ is the layer spacing.
$B$ 
is the scattering matrix, 
\begin{equation}
 B = \begin{bmatrix} 0 & 0 \\ 0 & sl \end{bmatrix},
 \label{eq:Billoir:scattering}
\end{equation}
where
$s \equiv \left(\gfrac{p_0}{p}\right)^2 \gfrac{\Delta x}{l X_0}$ is the average
multiple-scattering angle variance per unit track length,
$\theta_0^2 = s \times l$.
$M$ is the measurement matrix,
\begin{equation}
 M = \begin{bmatrix} \imath l & 0 \\ 0 & 0 \end{bmatrix},
 \label{eq:Billoir:measurement}
\end{equation}
where
$\imath \equiv\gfrac{N+5}{L\sigma^2}\approx\gfrac{1}{l\sigma^2}$
is the information density per unit track length,
 $L = N \times l$ is the full detector thickness.
 %

Billoir considers the two particular cases of ``scatters at one
point'' (detector layers separated by an empty space)
that we name here a segmented detector
and ``uniformly
distributed scattering'' (homogeneous detector) \cite{Billoir:1983mz}.
These concepts are defined more precisely below, following Innes
\cite{Innes:1992ge}.

\subsection{Segmented detector}
\label{sub:sec:segmented:detector}

Expressing $I_n$ as $I_n = A_nB_n^{-1}$ (\cite{Grewal:2001}, page 149)
we obtain
\begin{equation}\label{eq:fonctio}
\begin{bmatrix}
A_{n+1} \\ B_{n+1}
\end{bmatrix} = \begin{bmatrix}
D^T+MD^{-1}B & MD^{-1} \\ D^{-1}B & D^{-1}
\end{bmatrix}\begin{bmatrix}
A_{n} \\ B_{n}
\end{bmatrix},
\end{equation}
and
\begin{equation}\label{eq:In}
 I_{n+1} = A_{n+1}B_{n+1}^{-1} .
\end{equation}

Noting
\begin{equation}\label{eq:def:Phi}
 \Phi \equiv \begin{bmatrix}
 D^T+MD^{-1}B & MD^{-1} \\ D^{-1}B & D^{-1}
 \end{bmatrix}
 ,
\end{equation}
we obtain 
\begin{equation}\label{eq:discpuiss}
 \begin{bmatrix}
 A_{n} \\ B_{n}
 \end{bmatrix} = \Phi^n\begin{bmatrix}
 A_{0} \\ B_{0}
 \end{bmatrix}.
\end{equation}
 
$A_n$ and $B_n$ are obtained from the eigenvalues of $\Phi$.
The covariance matrix becomes $V_n = B_nA_n^{-1} $.
We initialise the recurrence with $ A_0 = I_0 = 0$, $ B_0 = 1$. 
If $\begin{bmatrix} A_{n} \\ B_{n} \end{bmatrix}$ is a
solution of eq.~(\ref{eq:In}), then for any $\beta >0$,
\begin{equation}\label{eq:AB:Norm:beta}
 \gfrac{1}{\beta^n}\begin{bmatrix}
 A_{n} \\ B_{n}
 \end{bmatrix} = \gfrac{1}{\beta^n}\Phi^n\begin{bmatrix}
 A_{0} \\ B_{0}
 \end{bmatrix}
\end{equation}
 is a solution too.
Noting 
 $\tilde A_n = \gfrac {A_n}{\beta^n}$,
 $\tilde B_n = \gfrac {B_n}{\beta^n}$,
 $\tilde \Phi = \gfrac \Phi\beta$,
 we obtain 
\begin{equation}\label{eq:AB:afer:Diag} \begin{bmatrix}
 \tilde A_{n} \\ \tilde B_{n}
 \end{bmatrix} = \tilde \Phi^n\begin{bmatrix}
 \tilde A_{0} \\ \tilde B_{0}
 \end{bmatrix},
\end{equation}
 with $V_n = \tilde B_n\tilde A_n^{-1}$.
 $\Phi$ is found to satisfy 
 \begin{equation}
 \Phi^TJ\Phi=J,
 \end{equation}
with
\begin{equation}\label{eq:def:J}
J=\begin{bmatrix}0 & 0 & 1 & 0 & \\0 & 0 & 0 & 1\\-1 & 0 & 0 & 0\\0 & -1 & 0 & 0\end{bmatrix},
\end{equation}
from which $\Phi$ is a symplectic matrix.
General theorems enable a classification of $\Phi$ eigenvalues into
two ``invert'' and ``conjugate'' blocks, respectively (eq. (10) of
\cite{Adams:2015})
 \begin{equation}\label{eq:Phi:eigenspectrum}
 \left\lbrace \alpha, \alpha^*, \gfrac{1}{\alpha},\gfrac{1}{\alpha^*}\right\rbrace,
\end{equation}
 where ``$*$'' denotes complex conjugation.
 We choose $\alpha$ to have a norm larger than unity, $|\alpha|>1$.
We obtain \cite{math}
\begin{equation}\label{eq:alpha:of:x}
\alpha(x) = \frac{1}{2} j x^2+\frac{1}{2} \left(-x^4+4 j x^2\right)^{\frac{1}{2}}+1 ,
\end{equation}
where $j$ is the imaginary unit
and
$x \equiv \gfrac{l}{\lambda}$ is
the detector longitudinal sampling normalized to the detector
scattering length at momentum $p$ \cite{Innes:1992ge}:
\begin{equation}\label{eq:def:lambda}
 \lambda \equiv \gfrac{1}{\sqrt[4]{\imath s}}.
\end{equation}

An exploration of the consequences of a variation of the
initialisation of the recurrence parameters shows that the system
converges to the same solution $\alpha(x)$ regardless of the values of
$A_0, B_0$.
$B_0 \ne 0$ is needed so that $I_0$ is defined.
With $B_0=1$, $I_0 = A_0 = 0$ simply assumes that no {\sl a priori}
information is known about the track.

\begin{figure}[h]
 \centering 
 \includegraphics[width=0.645\linewidth]{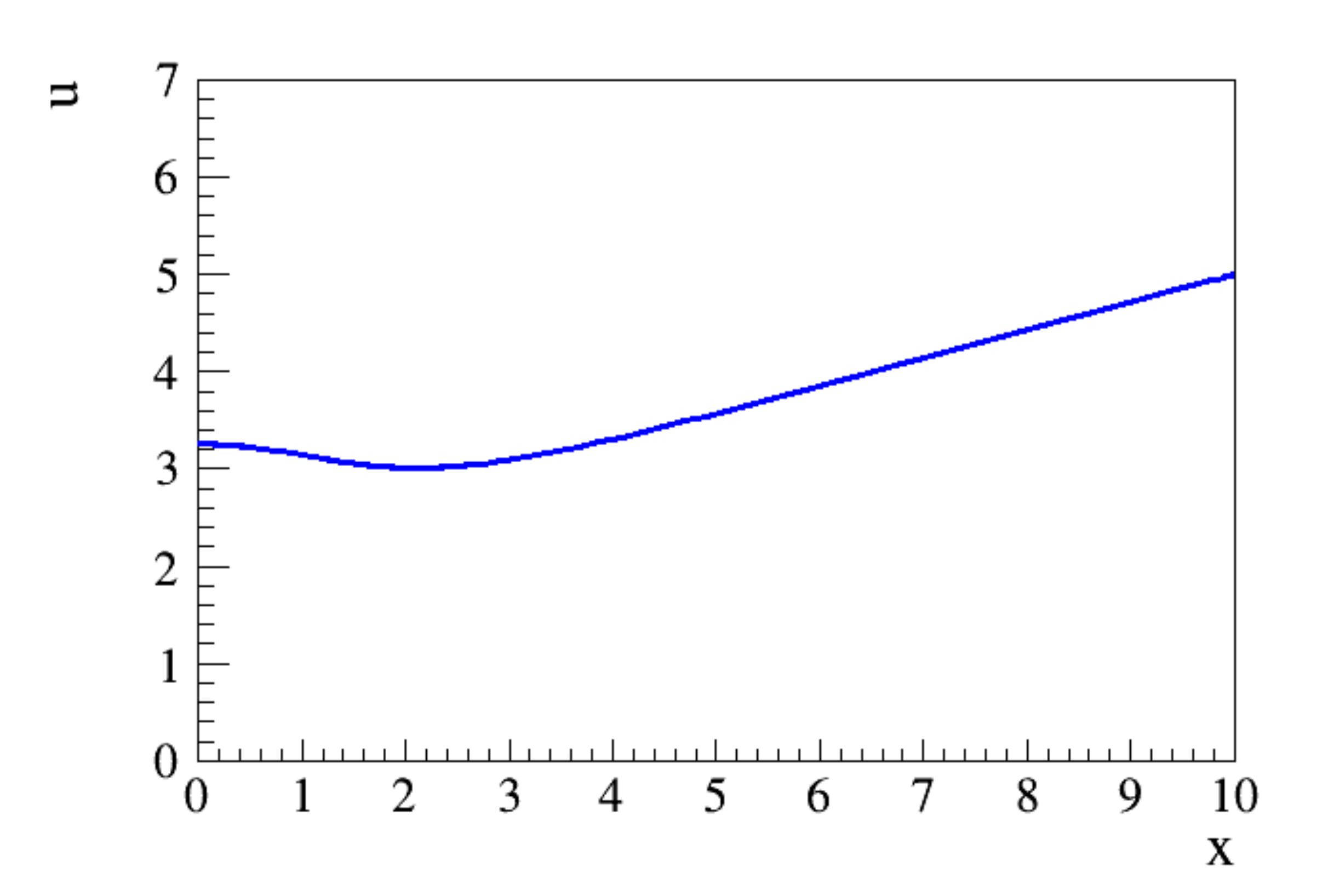} 
 \caption{
Value of $u$ for which
$\gfrac{1}{|\alpha|^{2 n}} < 10^{-4}$
as a function of $x$ (from eq. (\ref{eq:alpha:of:x})).
For $x<5$, convergence is reached for a detector thickness of
$4\lambda$. \label{fig:liminf}}
\end{figure}

We study the convergence of the covariance matrix while the Billoir
mechanism is in progress along the track (increasing $n$)
by setting $\beta = |\alpha|$, that is, 
 $\tilde \Phi = \gfrac{1}{|\alpha|}\Phi$.
$\tilde \Phi$ has two eigenvalues with modulus unity and two
eigenvalues with modulus $\gfrac{1}{|\alpha|^2}$.
The unity-modulus eigenvalues could be a major nuisance in the
behaviour of $I_n$ as a function of $n$, but when applying the Billoir
mechanism we observe that for some reason the amplitude the so-induced
oscillating terms is zero.
The convergence behaviour is then driven by the two other eigenvalues,
that is,
by terms proportional to $\gfrac{1}{|\alpha|^{2 n}}$.
That exponential convergence is illustrated in Fig. \ref{fig:liminf}
that shows the value of the detector thickness normalized to the
detector scattering length, $u$ \cite{Innes:1992ge}, 
\begin{equation}\label{eq:def:u}
 u \equiv \gfrac{L}{\lambda},
\end{equation}
for which
$\gfrac{1}{|\alpha|^{2 n}} < 10^{-4}$, as a function of $x$.

Note that the homogeneousness parameter $x$ and the thickness
parameter $u$ have a similar dependence on track momentum $p$, as
$u = x \times N$.

\subsubsection{Segmented detector: Thick detector limit}
\label{sub:sec:thick:detector}
 
The asymptotic expression at high $n$, that is, at high $u$ (thick
detector) is reached after the Billoir mechanism
(eq.~(\ref{eq:Billoir})) has converged: we obtain the discrete Riccati
equation:
\begin{equation}
I= D^T \left(I^{-1} + B \right)^{-1} D + M.
 \label{eq:Riccati}
\end{equation}

When the geometric, the multiple scattering and the measurement properties of
the detector are uniform (at least piecewise) the dynamics of the
particle is described by a time-invariant system and
eq.~(\ref{eq:Riccati}) is referred to as the ``algebraic'' Riccati
equation (DARE).
Equation (\ref{eq:Riccati}) has four solutions, but the fact that the
asymptotically stable solution must be positive definite (Theorem 2.2
of \cite{Adams:2015}) leaves us with only one.

\paragraph{Segmented detector: Exact solution} ~

We obtain \cite{math}
\begin{equation}
 V=\left(
 \begin{array}{cc}
 \frac{4 l^3 s}{ x^3 \left( 2 x + \sqrt{x^2 + 4 j} - \sqrt{x^2 - 4 j}\right)}
 &
 \frac{s l^2 \left(\sqrt{x^2 + 4 j} + \sqrt{x^2 - 4 j}\right)}{x^2 \left(\sqrt{-x^2 - 4 j} - \sqrt{4 j - x^2} -2 j x\right)}
 \\
 -\frac{4 l^2 s}{x^2 \left(\sqrt{-x^2 - 4 j} - j x \right) \left(j x + \sqrt{4 j - x^2}\right)}
 &
 \frac{2 l s \left(\sqrt{x^2 + 4 j} + \sqrt{x^2 - 4 j}\right)}{\left(x + \sqrt{x^2 + 4 j}\right) \left(j x + \sqrt{4 j - x^2}\right)}
 \\
 \end{array}
 \right)
. \label{eq:exact:discret} 
 \end{equation}

Even though it is not explicit from eq. (\ref{eq:exact:discret}), $V$ is
found to be a real matrix, which is decent for a covariance matrix.

\paragraph{Segmented thick detector: Small $x$ behaviour: Homogeneous detector limit} ~

\begin{figure}[h]
 \centering 
 \includegraphics[width=0.49\linewidth]{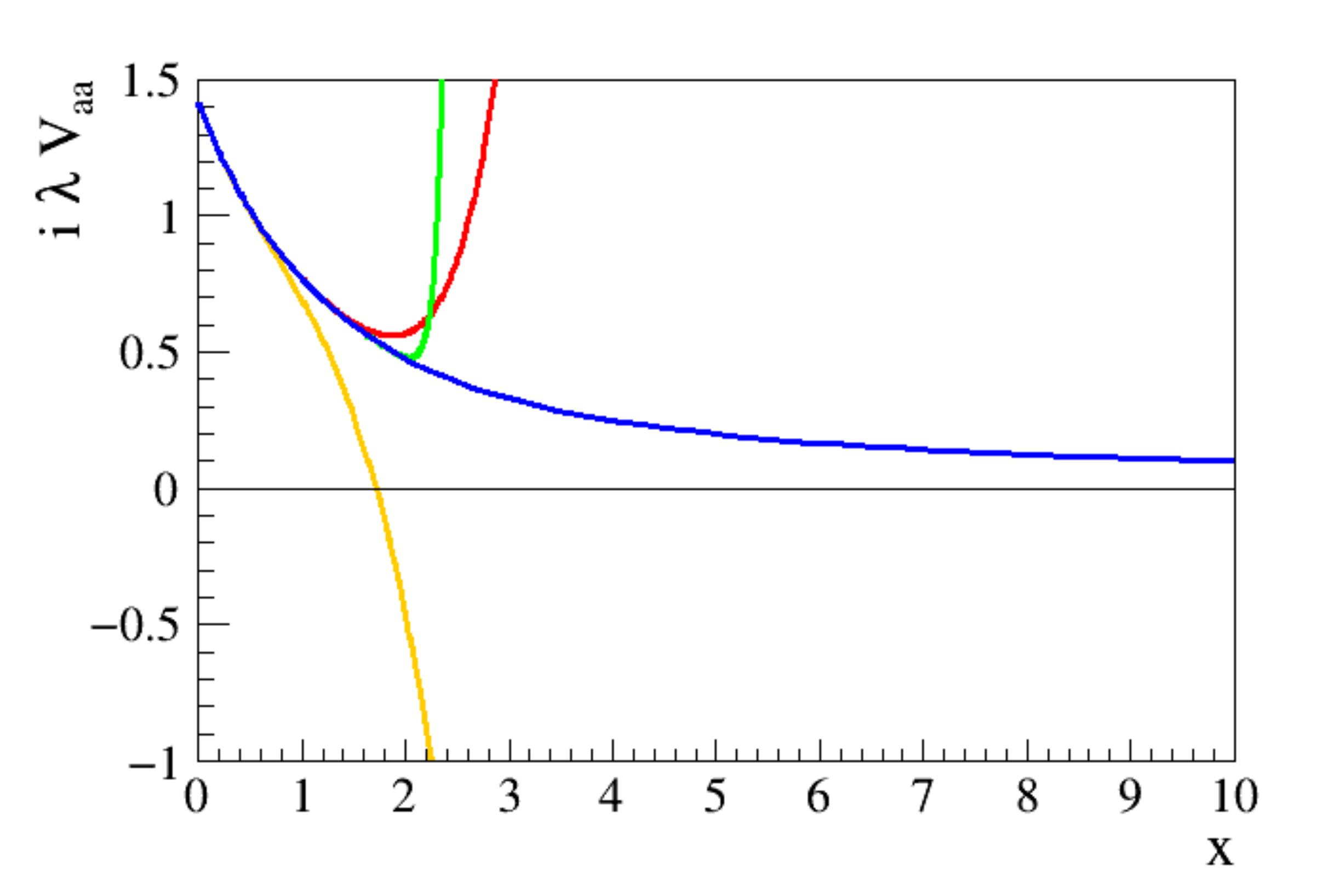} 
\hfill
 \includegraphics[width=0.49\linewidth]{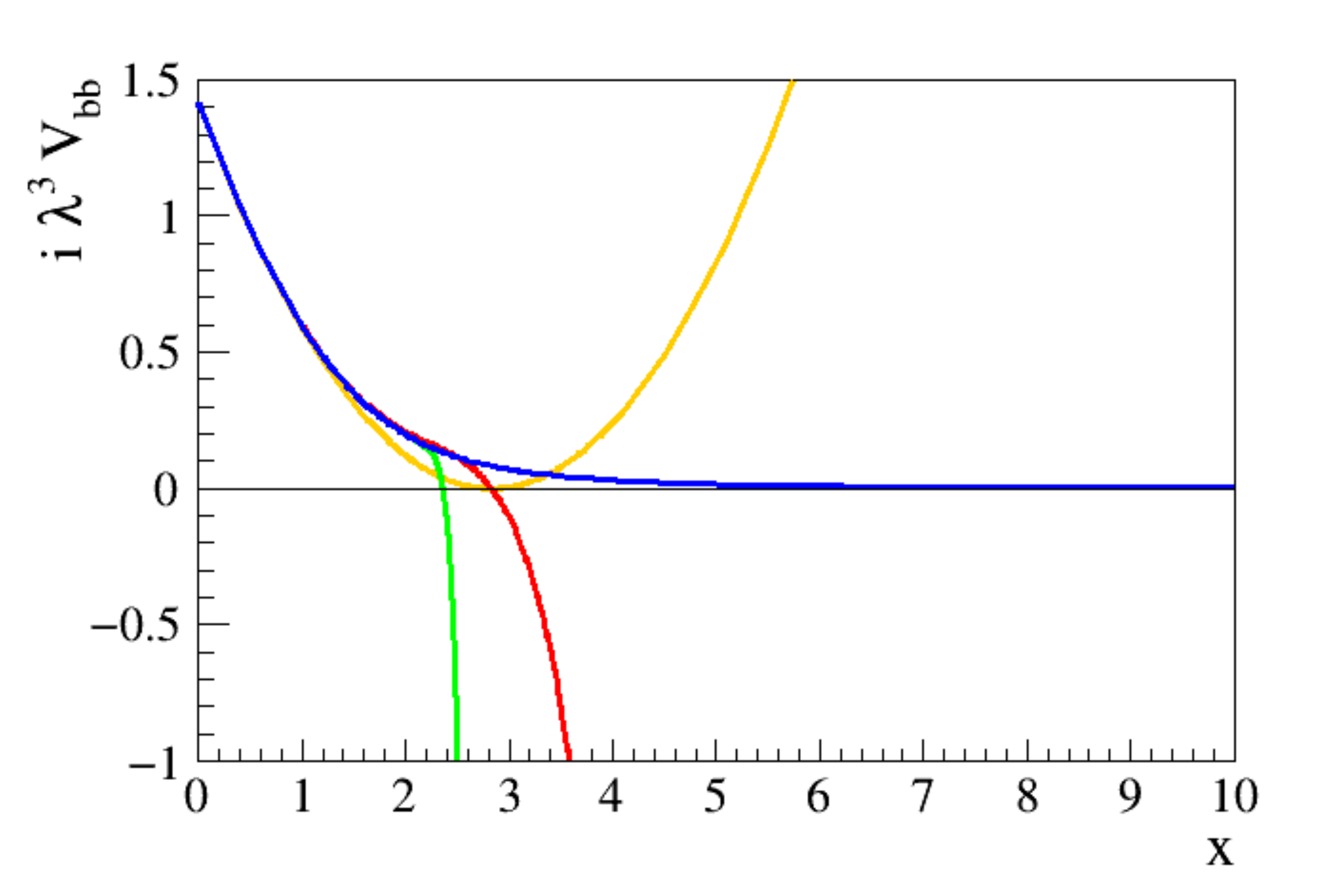}
 \put (-95,25) {$6$}
 \put (-85,90) {$2$}
 \put (-130,25) {$30$}
 \put (-60,60) {$es$}
 \put (-285,25) {$3$}
 \put (-275,90) {$6$}
 \put (-300,90) {$30$}
 \put (-210,60) {$es$}
 \caption{
Thick detector: $\imath\lambda V_{aa}$ and $\imath\lambda^3V_{bb}$ as a function of $x$.
Comparison of the Taylor expansions to several orders, 
 eqs. (\ref{eq:2Ddis1}) and (\ref{eq:2Ddis2}) 
to the exact solution ``es'', eq. (\ref{eq:exact:discret}). 
 \label{fig:vdisc}}
\end{figure}
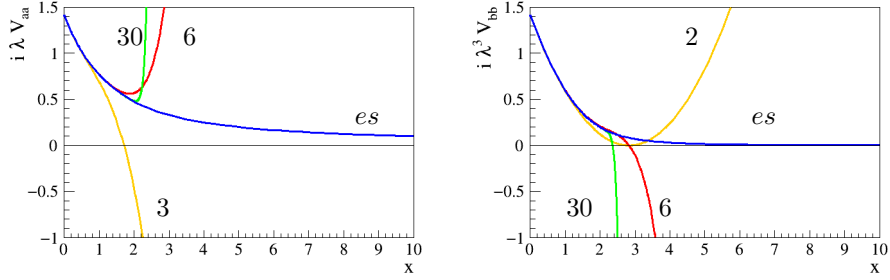

The Taylor expansion close to $x=0$ is found to be 
\begin{equation}\label{eq:2Ddis1}
 V_{aa}=\frac{\sqrt2}{\imath\lambda}\left[1-\frac x{\sqrt2}+\frac{3}8x^2-\frac{\sqrt2}8x^3+\frac{9}{128}x^4+O[x^5]\right],
\end{equation}
 
\begin{equation}\label{eq:2Ddis2}
 V_{bb}=\frac{\sqrt2}{\imath\lambda^3}\left[1-\frac x{\sqrt2}+\gfrac{1}{8}x^2+\frac{x^3}{128}-\frac{1}{1024}x^4+O[x^5]\right].
\end{equation}
These expressions are similar\footnote{We have detected a misprint,
 though: the factor $-5/8$ in their expressions of $V_{aa}$ is here
 corrected to $3/8$.}
to what was found by Billoir (\cite{Billoir:1983mz}, p364, no
 magnetic field) and
Innes (\cite{Innes:1992ge}, eq.~(8), magnetic field).
The Taylor expansion is found to converge for $x \lesssim 2$ 
(Fig. \ref{fig:vdisc}).

\paragraph{Segmented thick detector: Large $x$ behaviour: Coarse segmentation limit} ~

\begin{figure}[h]
\centering 
 \includegraphics[width=0.49\linewidth]{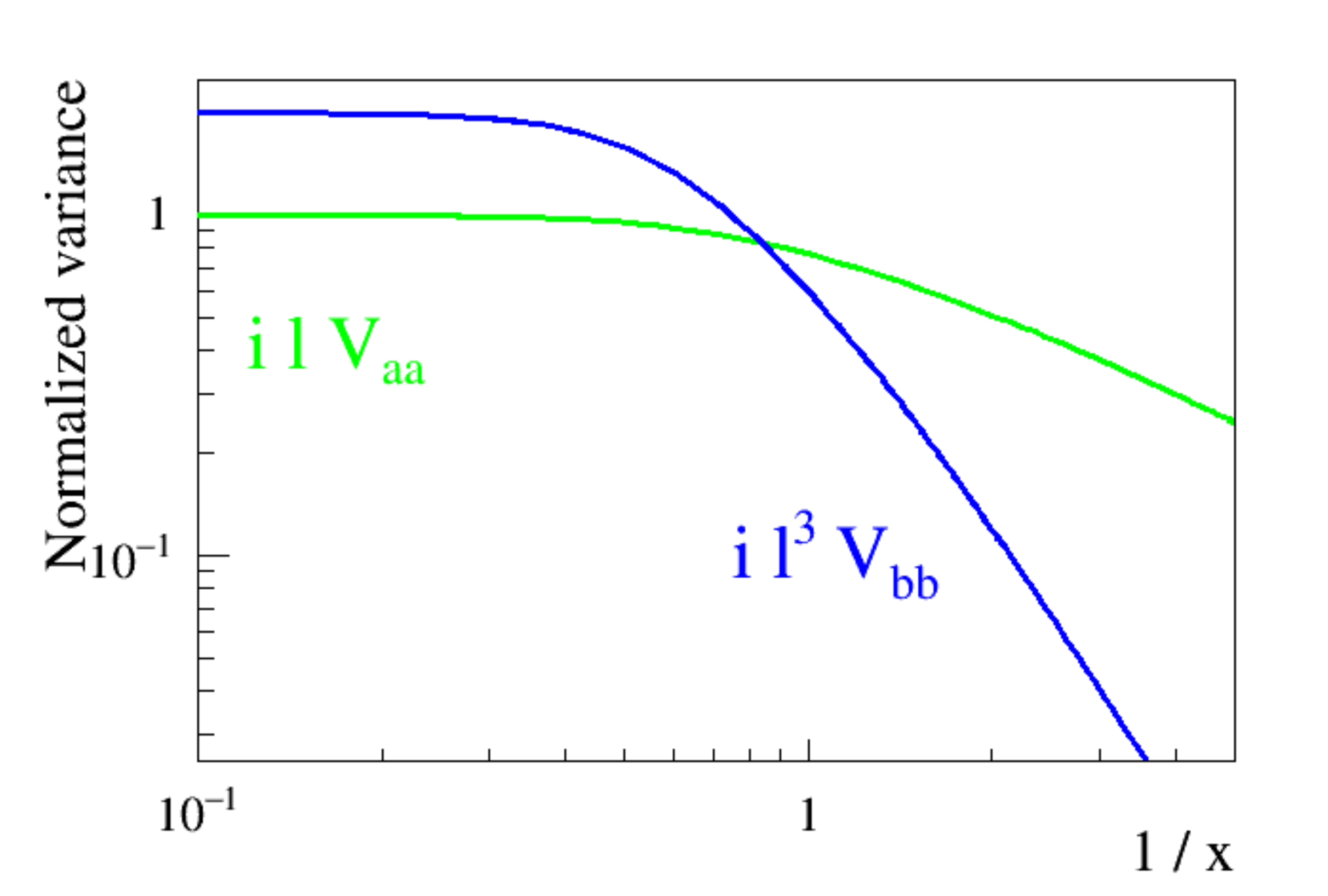} 
\caption{\label{g:lambda}
 Thick detector: Normalized variance
 $\imath l V_{aa}$ and $\imath l^3 V_{bb}$ as a function of $1/x$
 (eq. (\ref{eq:exact:discret})).}
\end{figure}

The asymptotic behaviour 
 of the coarsely instrumented detector (high $x$) 
 is presented in Fig. \ref{g:lambda}.

For $1/x = 0$ we obtain $\imath l V_{aa}=1$ and $\imath l^3 V_{bb}=2$,
that is the obvious
\begin{equation}
 V_{aa} = \sigma^2,
 \ \ \ 
 V_{bb} =2 \left(\gfrac{\sigma}{l}\right)^2
 :
 \label{eq:coarse}
\end{equation}
the scattering is so intense that the intercept (angle) measurement is
based on the first (two first) layer(s), respectively.
A thick coarse detector can be defined by $1 / x < 0.5$, that is, $l > 2 \lambda$
(Fig. \ref{g:lambda}).
The $\gfrac{1}{x}$ Taylor expansion is:
\begin{eqnarray}
 V_{aa}&=&\frac1{\imath l}\left[1-\frac1{x^4}+O\left(\frac1{x^8}\right)\right]
 , \\
 V_{bb}&=&\frac1{\imath l^3}\left[2-\frac{10}{x^4}+O\left(\frac1{x^8}\right)\right]
 .
\end{eqnarray}

\subsection{Homogeneous Detector}
\label{sub:sec:homogeneous:detector}

\begin{figure}[h]
 \centering 
\includegraphics[width=0.49\linewidth]{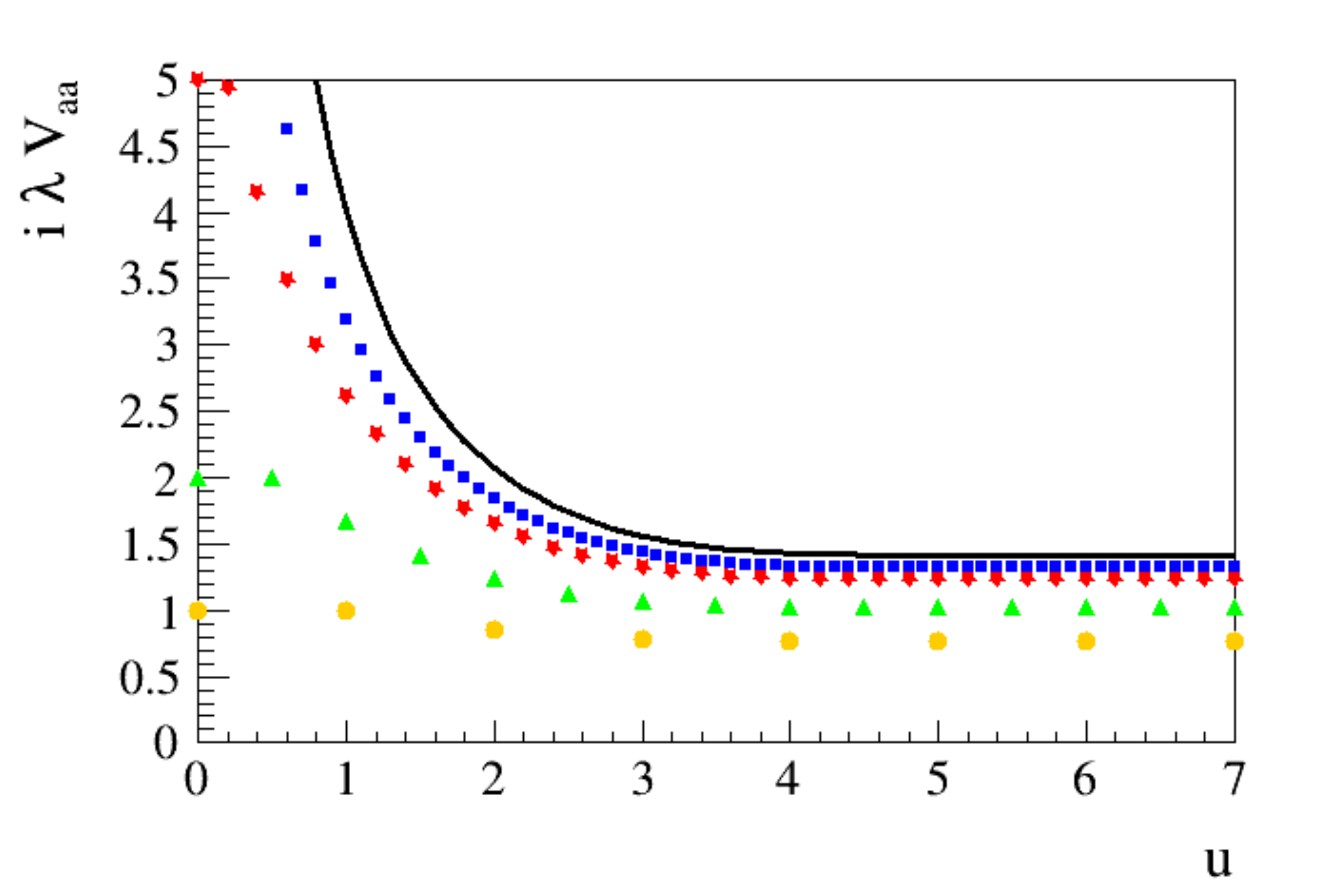}
\hfill
\includegraphics[width=0.49\linewidth]{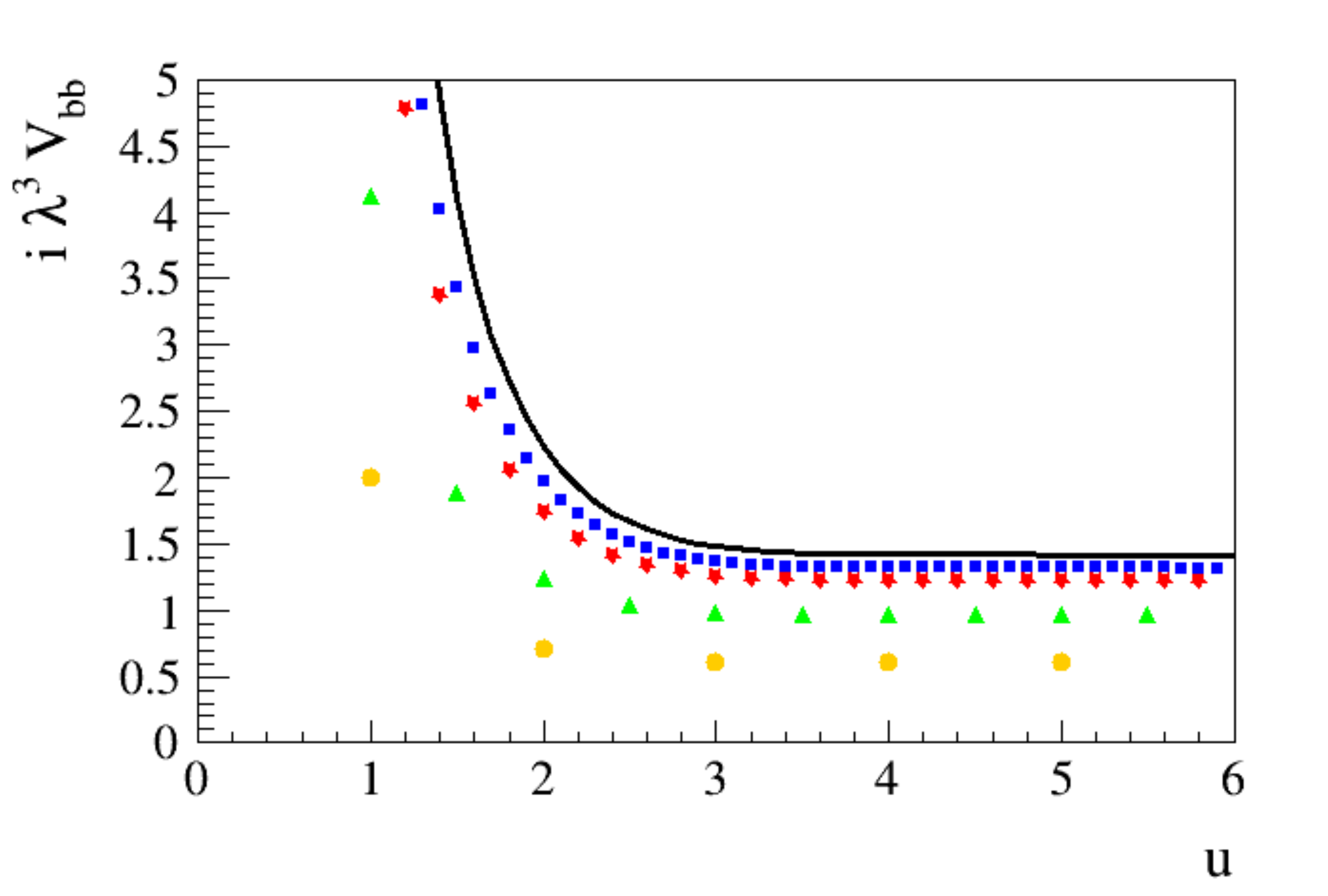}

\caption{Normalized covariance coefficients ${\imath \lambda V_{aa}}$ (left) and
 ${\imath \lambda^3 V_{bb}}$ (right) as a function of $u$ for various
 values of $x \in\{0,0.1,0.2,0.5,1\}$.
Curve, $x=0$ (eq. (\ref{eq:CARE:exact}));
 squares, $x=0.1$;
 stars, $x=0.2$;
 triangles, $x=0.5$;
 bullets, $x=1.0$.
In both cases ($a$ and $b$), $x \lesssim 0.2$ is found to be a good
approximation of the homogeneous detector ($x=0$).
 \label{fig:V_x}}
\end{figure}

A homogeneous detector is described having $l$ tend to $0$ while $s$
and $\imath$ are kept constants.
Fig. \ref{fig:V_x} shows that for all values of $u$, the intercept and angle
variances become very close to the homogeneous
limit $(x=0)$ for $x\lesssim 0.2$.

From the discrete evolution equation, eq.~(\ref{eq:Billoir}),
and denoting $I_n=I(nl)=I(L)$, we obtain
\begin{equation}\label{eq:cont}
\dot I(L)=D'^TI(L)+I(L)D'-I(L)B'I(L)+M',
\end{equation}
where the dot denotes the derivation with respect to $L$ and 
with
\begin{equation}\label{eq:def:matrix:prime}
 D'=\begin{bmatrix} 0 & 1 \\ 0 & 0\end{bmatrix},
\ \ \ 
 B'=\begin{bmatrix} 0 & 0 \\ 0 & s \end{bmatrix},
\ \ \ 
 M'=\begin{bmatrix} \imath & 0 \\ 0 & 0 \end{bmatrix}.
\end{equation}

After convergence (thick detector), we obtain the continuous algebraic Riccati equation (CARE): 
\begin{equation}\label{eq:CARE}
D'^TI(L)+I(L)D'-I(L)B'I(L)+M' = 0,
\end{equation}

\paragraph{Homogeneous Detector: Small $u$ behaviour} ~

We first use Innes' method to compute an approximate solution. Attempting a Taylor expansion in $u$, 
$I(u)=\sum I^ku^k$, we obtain:
\begin{eqnarray} \label{eq:serie_entiere}
 I^0 & = & 0, \nonumber
 \\ 
 I^1 & = & M', \nonumber
 \\ 
 I^{k+1} & = & \frac{1}{k+1}\left(-\sum_{i=0}^kI^iB'I^{k-i}+D'^TI^k+I^kD'\right) .
\end{eqnarray}
 \begin{itemize}
\item
In our case ($B=0$, no curvature) we obtain: 
\begin{equation}\label{eq:2Dcontinu1}
V_{aa}=\frac{4}{\imath\lambda u}\left[1+\frac{u^4}{416}-\frac{127u^8}{15\,891\,876\,000}+O(u^{12})\right]
\end{equation}

\begin{equation}\label{eq:2Dcontinu2}
V_{bb}=\frac{12}{\imath\lambda^3u^3}\left[1+\frac{13u^4}{420}-\frac{13\,429u^8}{529\,200}+O(u^{12})\right]
\end{equation}
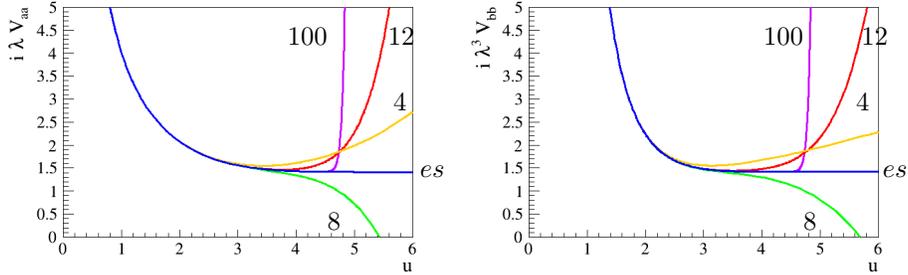
\begin{figure}[h]
 \centering
 \includegraphics[width=0.49\linewidth]{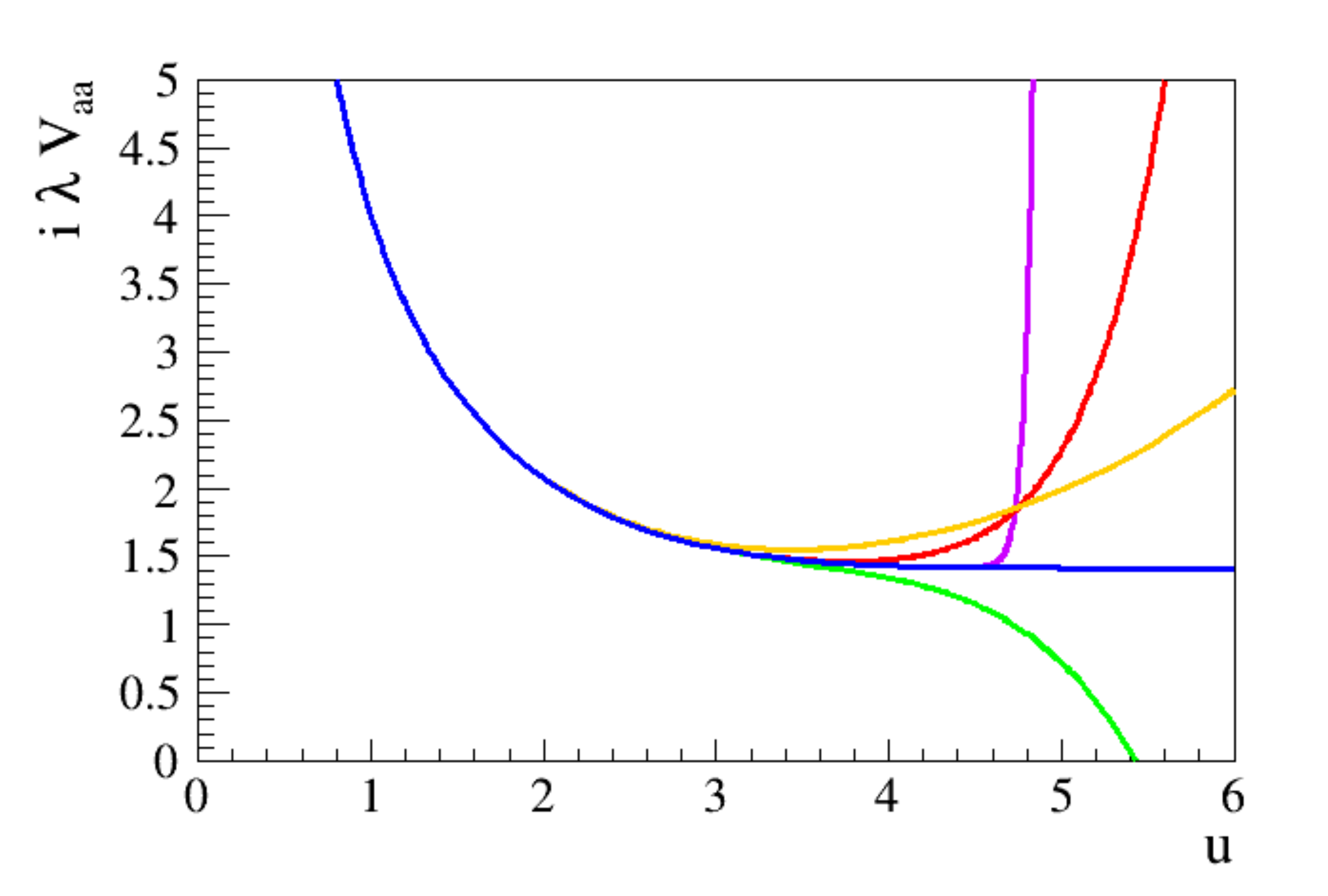} 
\hfill
\includegraphics[width=0.49\linewidth]{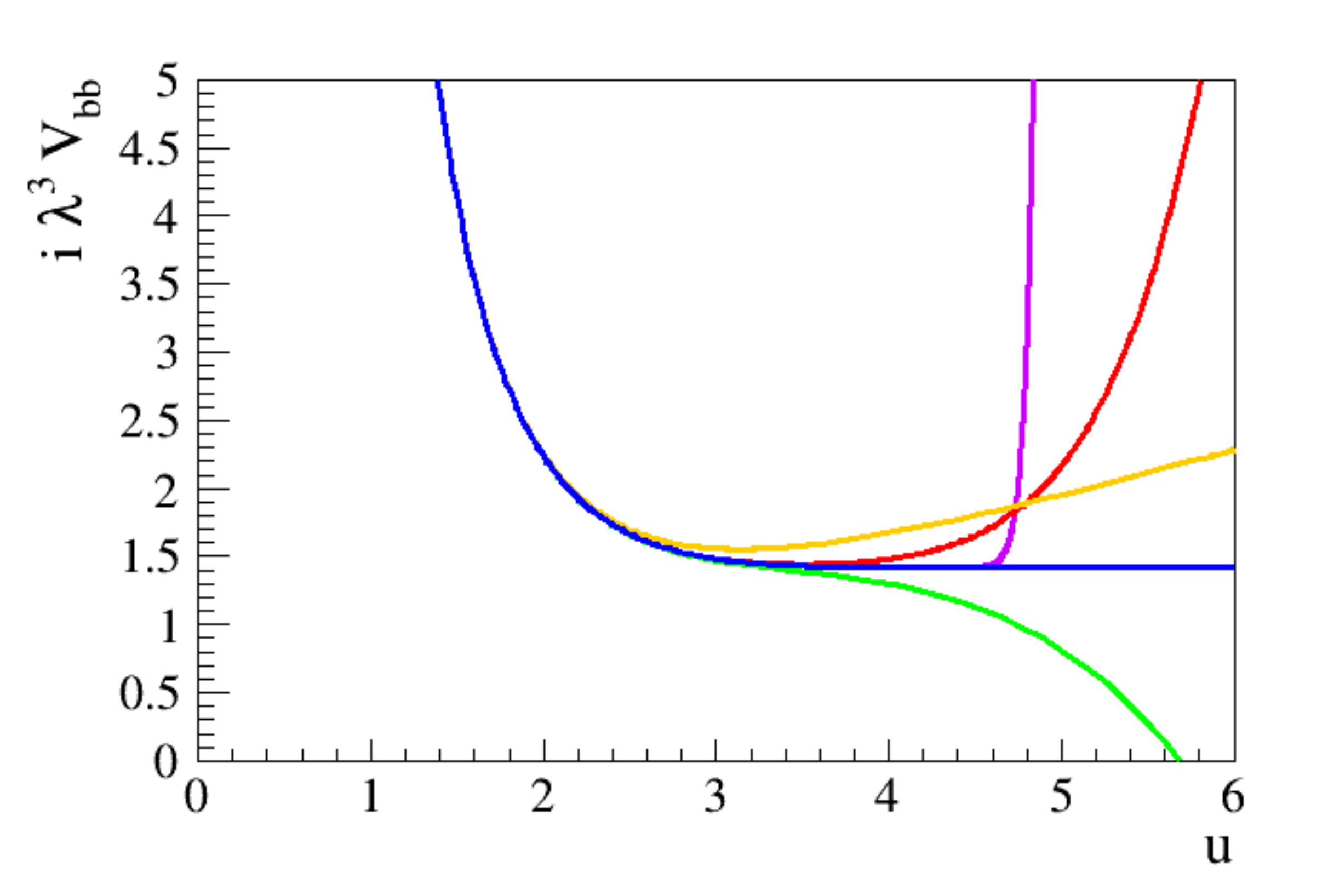}
 \put (-40,20) {$8$}
 \put (-55,90) {$100$}
 \put (-18,90) {$12$}
 \put (-20,65) {$4$}
 \put (-10,40) {$es$}
 \put (-220,20) {$8$}
 \put (-235,90) {$100$}
 \put (-197,90) {$12$}
 \put (-195,65) {$4$}
 \put (-185,40) {$es$}
 \caption{
 Homogeneous detector: $\imath\lambda V_{aa}$ and
 $\imath\lambda^3V_{bb}$ as a function of $u$.
 Comparison of the Taylor expansions to several orders,
 eqs. (\ref{eq:2Dcontinu1}) and (\ref{eq:2Dcontinu2}), 
 to the exact solution obtained from the resolution ``es'' of eqs.\,(\ref{eq:Phi:prime})-(\ref{eq:contexp}).
\label{fig:2DcDL}}
\end{figure}

\item
For $B \ne 0$ and a fit with curvature, we obtain the results of 
Innes (eq.~(9) of \cite{Innes:1992ge}).
\end{itemize}
These Taylor expansions converge for
$u \lesssim 3.5$ (no curvature, Fig. \ref{fig:2DcDL}) and
$u \lesssim 7.0$ (with curvature, \cite{Innes:1992ge}).
The thin detector ($u=0$) value of $V_{bb}$ for $B=0$ is found to be
smaller than that for $B \ne 0$ by a factor of 16, as was discussed in
the corrigendum of \cite{Bernard:2012uf}:
in a fit with curvature, the correlation between the curvature and the
angle at the end(s) of the track degrades the angular resolution
badly;
this lasts until $u\approx 1$ that is $L \approx \lambda$ (plot not shown), after which all is
flooded by multiple scattering anyway.

\paragraph{Homogeneous Detector: Large $u$ behaviour: Thick detector limit} ~

Searching for expressions that are valid at high $u$, we follow again
Innes and search a solution of the continuous equation for $V$ that is
similar to eq.~(\ref{eq:cont}) for $I$.
Here the Taylor expansion is searched in $1/u$.
Searching a solution parametrized as 
$V(u)=V_0+\gfrac{1}{u}V_1$, we obtain
\begin{equation}\label{eq:2Dcontinuinfini}
V_{aa}=\frac{\sqrt2}{\imath\lambda},
 \ \ \ \
V_{bb}=\frac{\sqrt2}{\imath\lambda^3}.
\end{equation}
These values agree with that of
eqs. (\ref{eq:2Ddis1}), (\ref{eq:2Ddis2}) for $x=0$.
The term proportional to $1/u$ that was present in the case with
curvature (eq.~(11) of \cite{Innes:1992ge}) cancels here, which is
related to the exponential convergence seen
on eq. \eqref{eq:CARE:exact} (see also Fig. \ref{g:WrapUp}).

\subsubsection{Homogeneous thick detector: Exact solution}
\label{sub:sub:sec:exact:solution}
 
We solve the continuous algebraic Riccati equation in a way similar to
the discrete case \cite{Grewal:2001}: Expressing
\begin{equation}\label{eq:Phi:prime}
 \Phi'=\begin{bmatrix}D' & M' \\ B' & -D'^T\end{bmatrix}
\end{equation}
and $I(L)=X(L)Y^{-1}(L)$, and taking $I(0)=0$, we obtain:
\begin{equation}\label{eq:contexp}
 \begin{pmatrix}X(L) \\ Y(L)\end{pmatrix}=\exp \left[L\Phi'\right]\begin{pmatrix}0 \\ 1\end{pmatrix}.
\end{equation}

$\Phi'$ satisfies 
\begin{equation}
 \Phi'^TJ\Phi'^{-1}=-J 
\end{equation}
and is a hamiltonian matrix (section (4.8) of \cite{Grewal:2001},
\cite{Yuantong}), which implies that
 $\exp \left[\Phi'\right]$ is a symplectic matrix and therefore that
$\Re\{\mathrm{Tr}(\Phi')\}=0$.
Furthermore all eigenvalues of $\Phi'$ are found to be non singular
 \cite{math}:
\begin{equation}\label{eq:Phi:prime:eigenspectrum}
 \mathrm{Spec}(\Phi')=\left\lbrace\frac{1}{\lambda} e^{\frac{-3j\pi}4},\frac{1}{\lambda} e^{\frac{-j\pi}4},-\frac{1}{\lambda} e^{\frac{-3j\pi}4},-\frac{1}{\lambda} e^{\frac{-j\pi}4}\right\rbrace
 .
\end{equation}

Solving eq. \eqref{eq:contexp} we obtain \cite{math}

\hspace{-2cm}
\begin{minipage}[b]{\linewidth}
\begin{equation}
V = 
\left(
\begin{array}{cc}
 \gfrac{\sqrt{2}}{\lambda \imath} \frac{\left(-j+e^{2 e^{\frac{j\pi}4} u}-e^{2 e^{\frac{3j\pi}4} u}+j e^{2 j \sqrt{2} u}\right)}{\left(1+e^{2 e^{\frac{j\pi}4} u}+e^{2 e^{\frac{3j\pi}4} u}-4 e^{j \sqrt{2} u}+e^{2 j \sqrt{2} u}\right)} & \gfrac{1}{\lambda^2 \imath} \frac{\left(-1+e^{2 e^{\frac{j\pi}4} u}\right) \left(-1+e^{2 e^{\frac{3j\pi}4} u}\right)}{\left(1+e^{2 e^{\frac{j\pi}4} u}+e^{2 e^{\frac{3j\pi}4} u}-4 e^{j \sqrt{2} u}+e^{2 j \sqrt{2} u}\right)} \\
 \gfrac{1}{\lambda^2 \imath} \frac{\left(-1+e^{2 e^{\frac{j\pi}4} u}\right) \left(-1+e^{2 e^{\frac{3j\pi}4} u}\right)}{\left(1+e^{2 e^{\frac{j\pi}4} u}+e^{2 e^{\frac{3j\pi}4} u}-4 e^{j \sqrt{2} u}+e^{2 j \sqrt{2} u}\right)} & \gfrac{\sqrt{2}}{\lambda^3 \imath} \frac{\left(j+e^{2 e^{\frac{j\pi}4} u}-e^{2 e^{\frac{3j\pi}4} u}-j e^{2 j \sqrt{2} u}\right)}{\left(1+e^{2 e^{\frac{j\pi}4} u}+e^{2 e^{\frac{3j\pi}4} u}-4 e^{j \sqrt{2} u}+e^{2 j \sqrt{2} u}\right)} \\
\end{array}
\right)
.\label{eq:CARE:exact}
\end{equation}
\end{minipage}

Even though it is not explicit from eq. (\ref{eq:CARE:exact}), $V$ is
found to be a real matrix, which is decent for a covariance matrix.
The convergence is driven by a term proportional to
$e^{-2ue^{\frac{j\pi}4}}$, which implies a convergence in
$e^{{-l\sqrt2} / \lambda}$.

\subsubsection{Variation of the angle variance along the track}
\label{sub:sub:sec:variation}

We have considered above the optimal measurement of the track
parameters at the vertex, $z=0$.
Here we examine the measurement at any
position along the track.
A track now consists of two segments (left and right), the fit
of each of which provides an estimate with its own covariance matrix.

\begin{figure}[h]
\centering 
\includegraphics[width=0.45\linewidth]{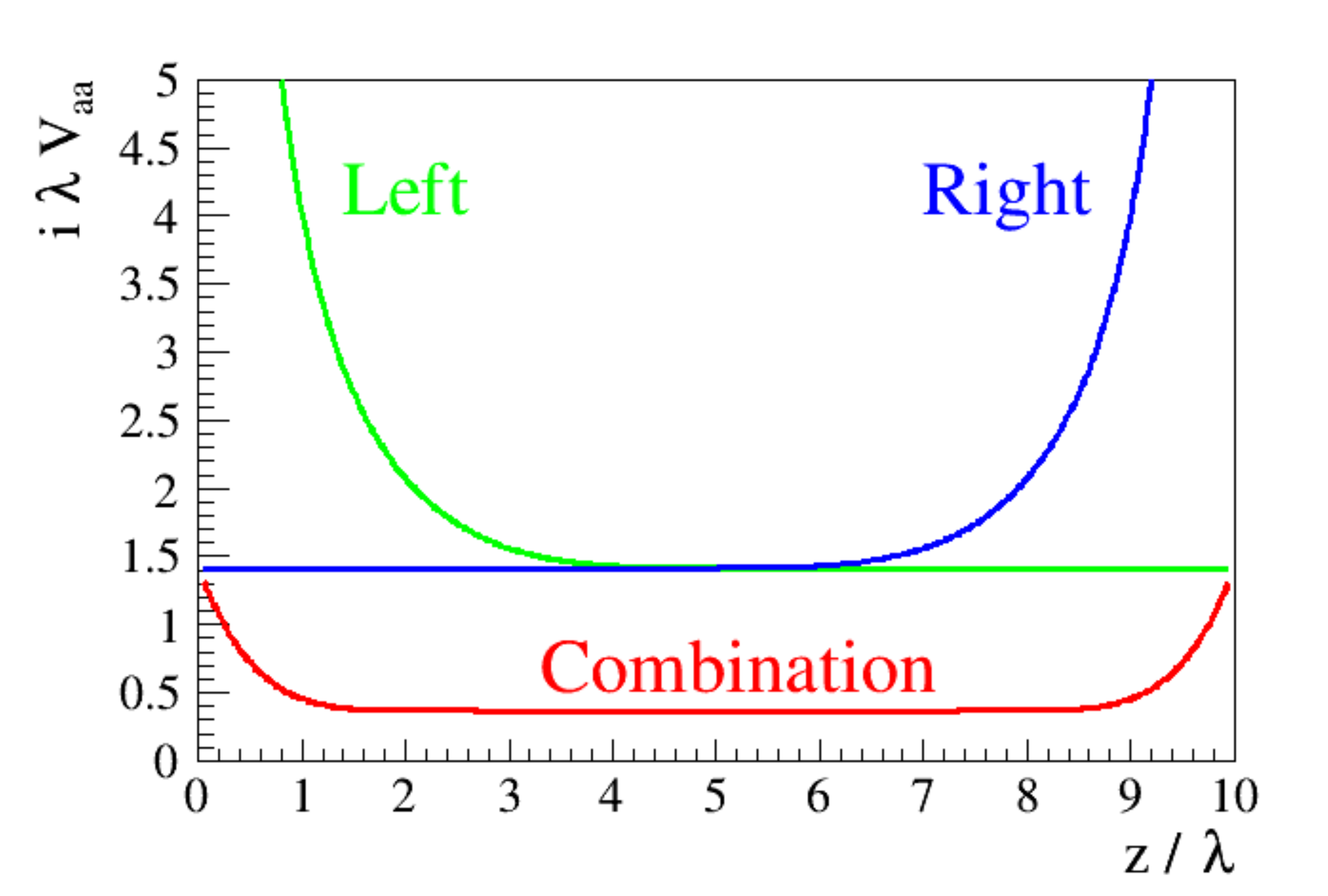}
\includegraphics[width=0.45\linewidth]{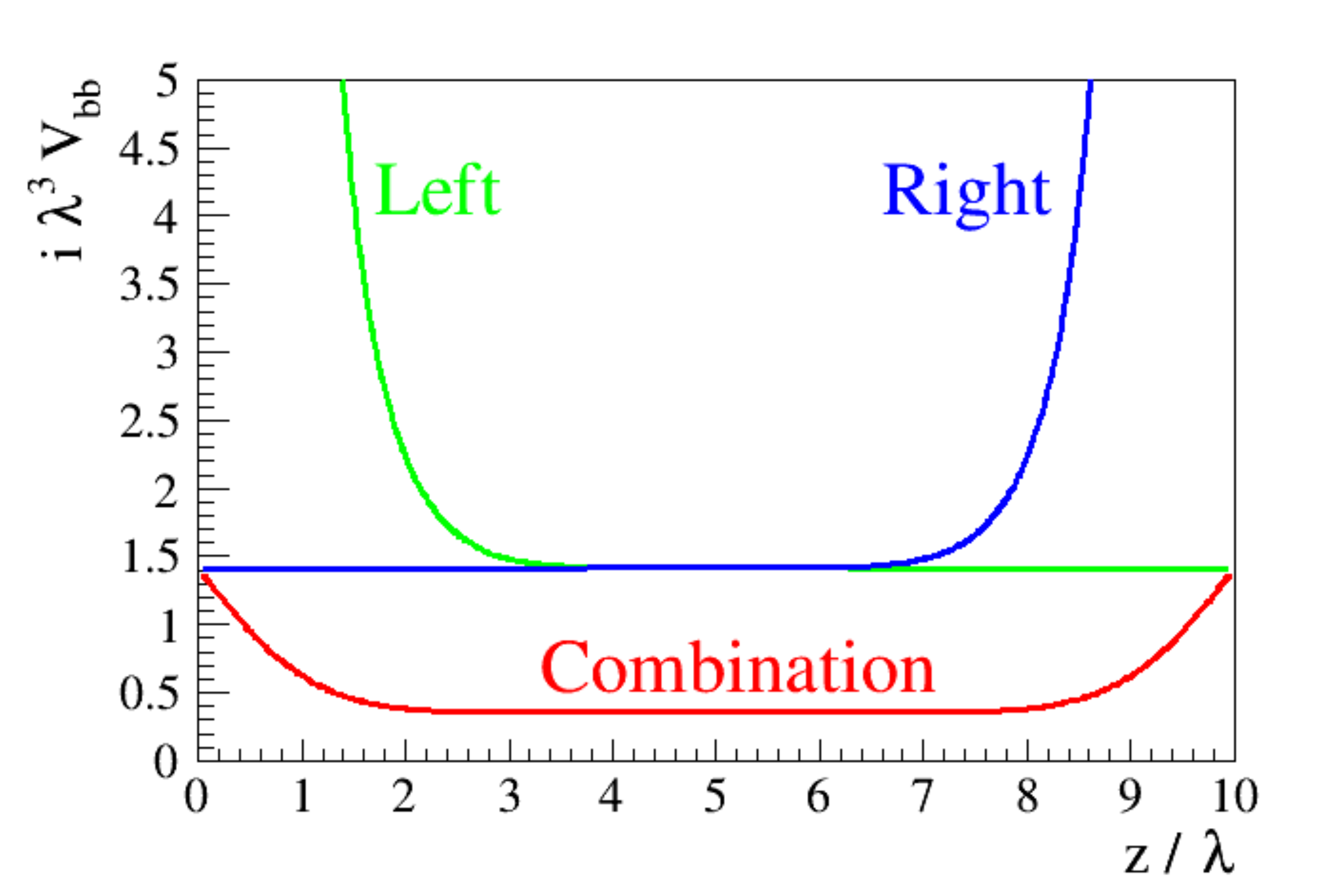}
\caption{\label{g:Vbb2Dopt}
Normalized variance along the track for an $u=10$ detector, as
estimated from the left and from the right side,
(eq. (\ref{eq:CARE:exact})) and of their optimal combination.
Left plot: intercept. Right plot: angle.
An improvement of a factor of 4 is visible on the plateau (i.e., far
from the track ends), that corresponds to a factor of 2 for the
standard deviation.}
\end{figure}

A first combination attempt is performed on the two variables ($a$ and $b$)
separately.
As is obvious, if the detector is thick on both sides, the two
estimates (right, left) of a track parameter (say: the angle) have the
same uncertainty on the plateau and their optimal combination
provides a gain in RMS precision of a factor of $\sqrt{2}$.
But that neglected the fact that in the combination, the other variable
should be constrained to have the same value on both sides too.
With a weighted variance estimation, 
a further gain of a factor of $\sqrt{2}$ is obtained on the plateau,
that is a total gain of a factor of $2$ with respect to individual
measurements (Fig. \ref{g:Vbb2Dopt}) as was observed experimentally by
running a KF on simulated tracks (Fig. 18 of \cite{Bernard:2013jea}).

\newtoggle{lignecontinue}
\toggletrue{lignecontinue}

\begin{figure}[h]
\centering 

\iftoggle{lignecontinue}{
\includegraphics[width=0.49\linewidth]{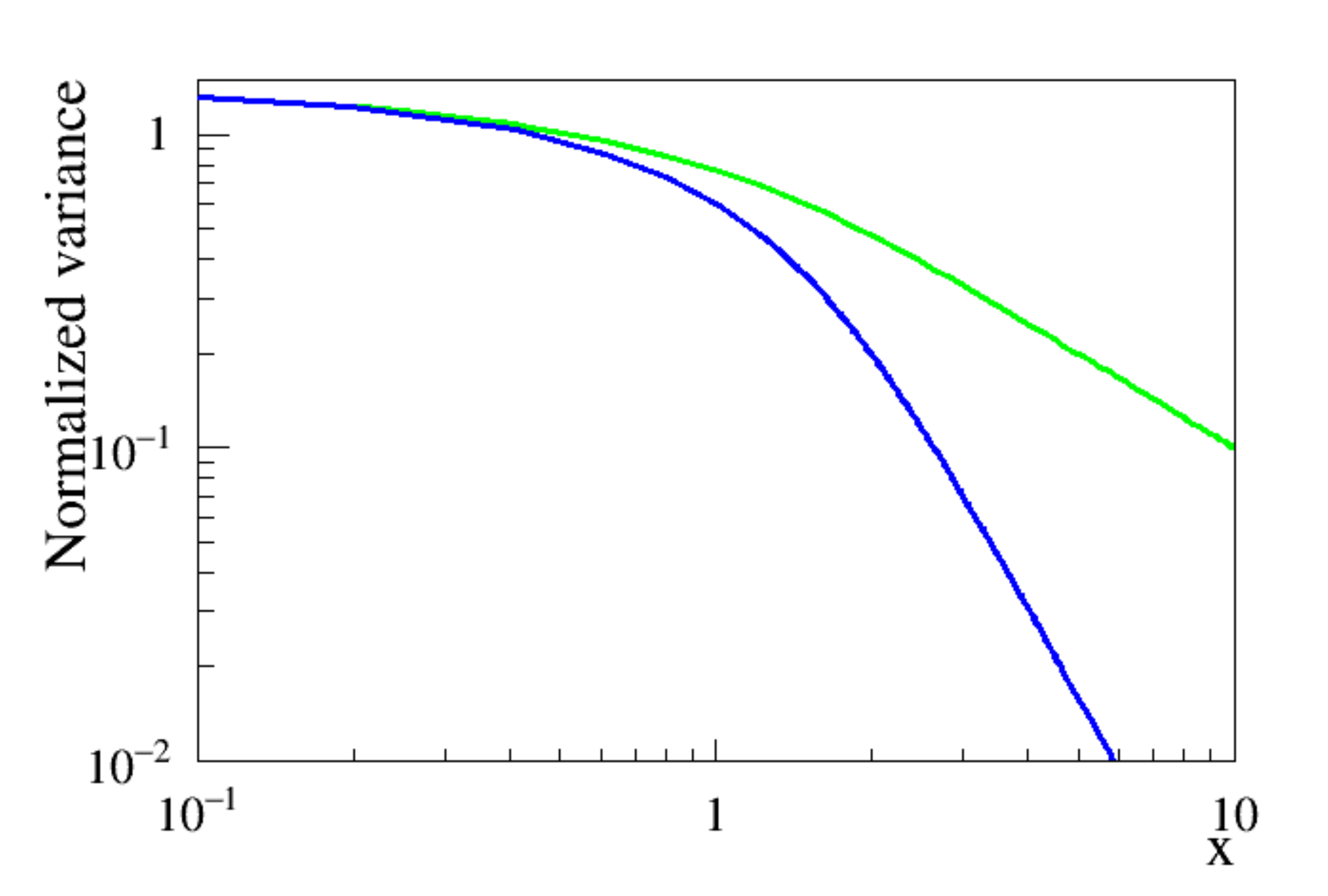}
}{
\includegraphics[width=0.49\linewidth]{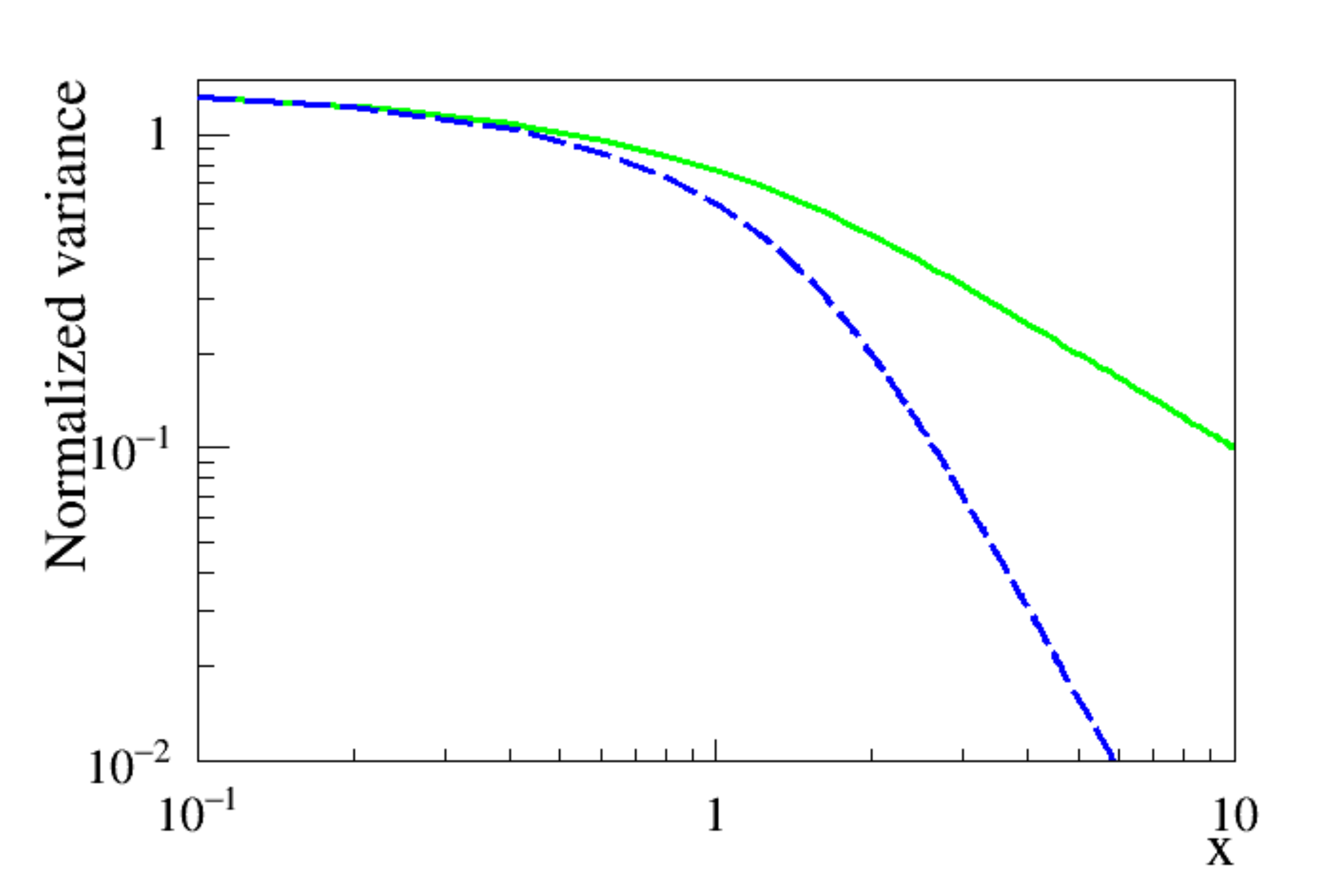}
}
\put(-140,50){\Blue{$\imath\lambda^3 V_{bb}$}}
\put(-140,70){\Green{$\imath\lambda V_{aa}$}}
\iftoggle{lignecontinue}{
\includegraphics[width=0.49\linewidth]{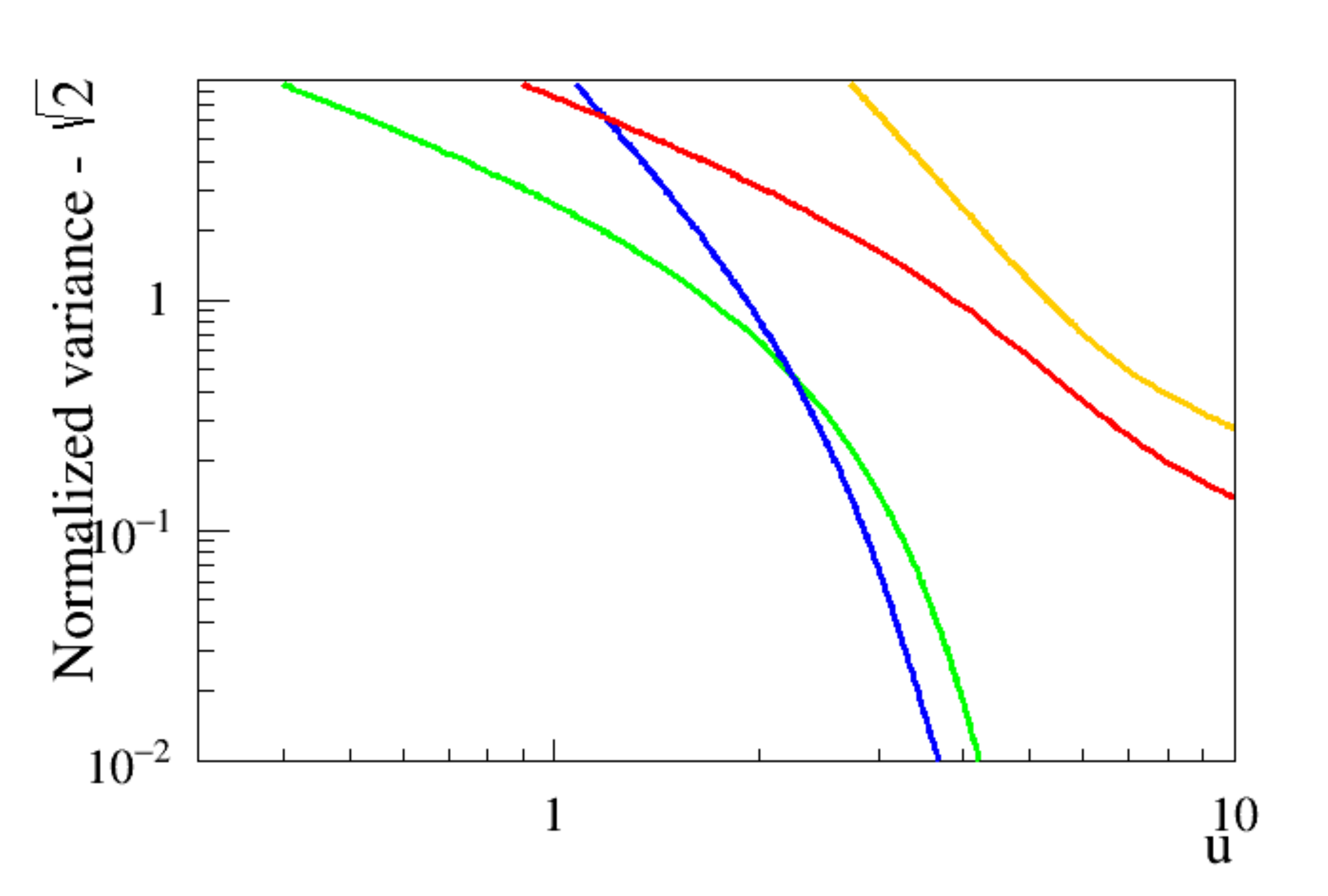}
}{
\includegraphics[width=0.49\linewidth]{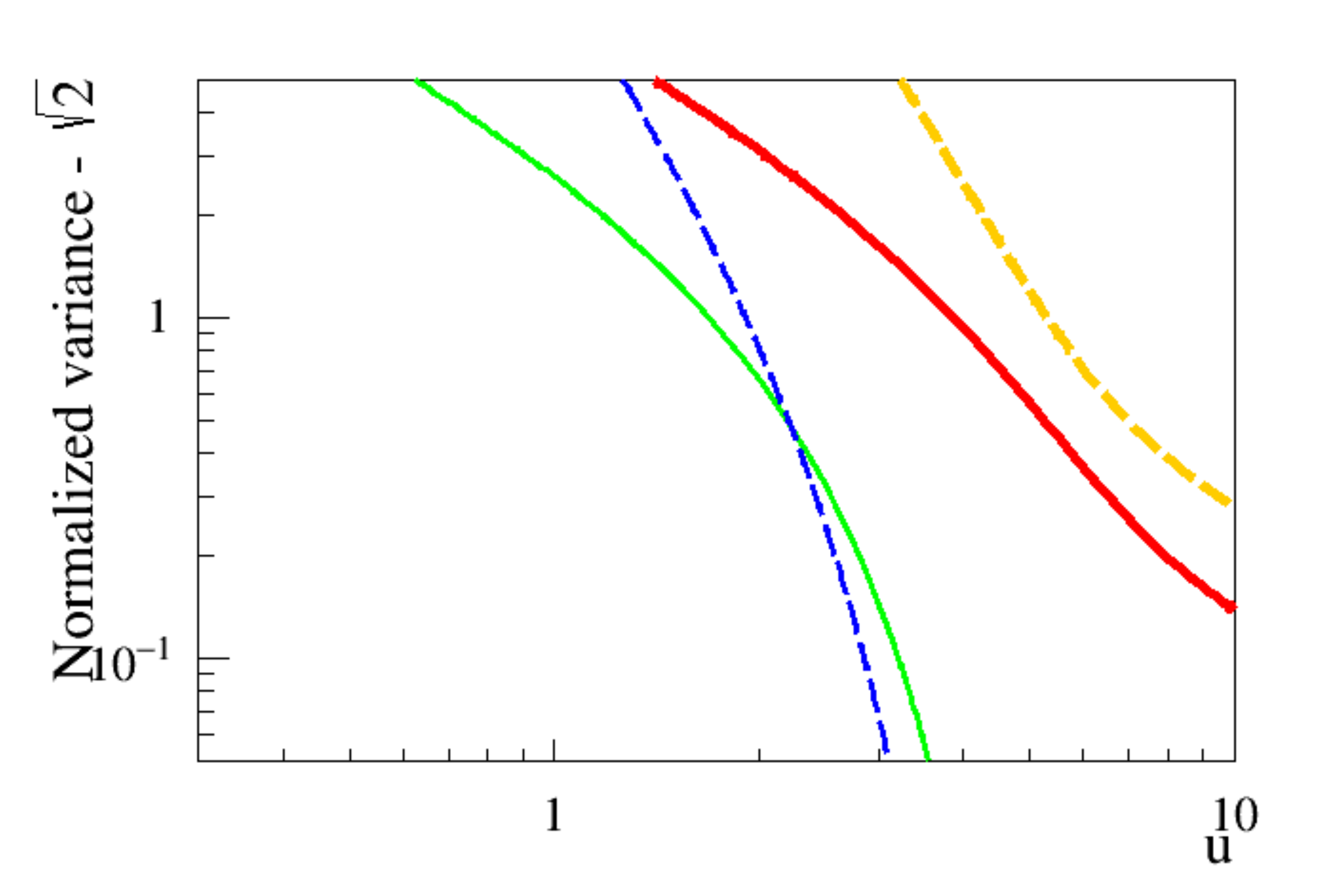}
}
\put(-140,60){\Blue{2D $\imath\lambda^3 V_{bb}$}}
\put(-140,78){\Green{2D $\imath\lambda V_{aa}$}}
\put(-140,22){\Orange{3D $\imath\lambda^3 V_{bb}$}}
\put(-140,40){\Red{3D $\imath\lambda V_{aa}$}}

\caption{\label{g:WrapUp} Exact solution wrap up:
Thick detector (left) and
homogeneous detector (right),
for 2D ($B = 0$, no curvature\iftoggle{lignecontinue}{}{, thin line}) and
3D ($B \ne 0$, curvature\iftoggle{lignecontinue}{}{, thick line}) configurations.
 Normalized variance 
 ($\imath\lambda V_{aa}$\iftoggle{lignecontinue}{}{, continuous line}, and
 $\imath\lambda^3V_{bb}$\iftoggle{lignecontinue}{}{, dashed line})
 minus the asymptote (either zero or $\sqrt{2}$),
 as a function of $x$ and of $u$, respectively.
 Note that for the thick detector (left plots), the 2D and 3D expressions are found to be the same.
 Curves are from
 eq. (\ref{eq:exact:discret}) 2D thick detector and
 eq. (\ref{eq:CARE:exact}) 2D homogeneous detector;
 the expressions in the 3D case are not shown in this paper.
}
\end{figure}

\subsection{Optimal tracking: wrap up}
\label{sub:sec:optimal:tracking:wrap:up}

We now build on the results of the previous subsections to obtain 
expressions of the variances in terms of the detector parameters.
We do so for segmented detectors.
The expressions for continuous detectors can be obtained with $\Delta x = l$.

\begin{center} \small
 \begin{tabular}{llll}
 $\imath \equiv $ &
 $ \gfrac{1}{l\sigma^2}$ &
 information density per unit track length
 \\
 ~
 \\
 $s \equiv $ &
 $ \left(\gfrac{p_0}{p}\right)^2 \gfrac{1}{X_0}\gfrac{\Delta x}{l}$ &
 average multiple-scattering angle variance per unit track length
 \\
 ~
 \\
 $ \lambda \equiv \gfrac{1}{\sqrt[4]{\imath s}} \approx $ &
 $ \sqrt{l \sigma \sqrt{\gfrac{X_0}{\Delta x}} \gfrac{p}{p_0}}$ &
 detector scattering length at momentum $p$
 \\
 ~
 \\
 $x \equiv \gfrac{l}{\lambda} \approx $ &
 $ \sqrt{\gfrac{l}{\sigma}\gfrac{p_0}{p}\sqrt{\gfrac{\Delta x}{X_0}}}$ &
 detector longitudinal sampling normalized to $ \lambda $
 \\
 ~
 \\
 $ u \equiv \gfrac{L}{\lambda} \approx $ &
 $ N\sqrt{\gfrac{l}{\sigma}\gfrac{p_0}{p}\sqrt{\gfrac{\Delta x}{X_0}}}$ &
 detector thickness normalized to $ \lambda $
\end{tabular}
\end{center}

With: 
\begin{itemize}
 \item small $x$ (large $p$), homogeneous detector (continuous equation),
 \item large $x$ (small $p$), segmented detector (discrete equation)
\end{itemize}
and
\begin{itemize}
 \item small $u$ (large $p$), thin detector,
 \item large $u$ (small $p$), thick detector.
\end{itemize}
The variances are found to be asymptotically:

\hspace{-2cm}
\footnotesize
\begin{tabular}{|l|l|ll|ll|l}
 \hline
& & homogeneous & & coarse & \\
& & $x < 0.2$ & & $x > 2$ & \\
 \hline
 & & & & & \\
 & thin & $\gfrac{4}{\imath\lambda u} = \gfrac{4 \sigma^2}{N}$ & eq. (\ref{eq:2Dcontinu1}) & & 
 \\
$a$ & & & & & 
 \\
 & thick & $\gfrac{\sqrt2}{\imath\lambda} = \sqrt{2}\left(\gfrac{p}{p_0}\right)^{-1/2} (l \sigma^3)^{1/2} \left(\gfrac{\Delta x}{X_0}\right)^{1/4}$ & eq. (\ref{eq:2Dcontinuinfini}) & $\gfrac{1}{\imath l} = \sigma^2$ & eq. (\ref{eq:coarse})
 \\
 & & & & & 
 \\
 \hline
 & & & & & 
 \\
 & thin & $\gfrac{12}{\imath\lambda^3u^3} = \gfrac{12 \sigma^2}{l^2 N^3}$ & eq. (\ref{eq:2Dcontinu2}) & & 
 \\
$b$ & & & & & 
 \\
 & thick & $\gfrac{\sqrt2}{\imath\lambda^3} = \sqrt{2} ~ \left(\gfrac{\sigma}{l}\right)^{1/2} \left(\gfrac{\Delta x}{X_0}\right)^{3/4} \left(\gfrac{p}{p_0}\right)^{-3/2} = \left(\gfrac{p}{p_1}\right)^{-3/2} $ & eq. (\ref{eq:2Dcontinuinfini}) & $\gfrac{2}{\imath l^3} =2 \left(\gfrac{\sigma}{l}\right)^2$ & eq. (\ref{eq:coarse})
 \\
 & & & & & 
 \\
 \hline
\end{tabular}
\normalsize

~

where $p_1$ is a momentum that characterises the tracking
angular-resolution properties of a detector affected by multiple
scattering \cite{Bernard:2013jea}
\begin{equation}
 p_1 = p_0
 \left(\gfrac{\Delta x}{X_0}\right)^{1/2} 
 \left( \gfrac{2 \sigma}{l} \right)^{1/3}.
 \label{eq:p_1}
\end{equation}

\begin{itemize}
 \item
The two $V_{aa}$ asymptotes cross for $u = u_{c,a} = 2\sqrt{2} \approx 2.83$;
 \item
The two $V_{bb}$ asymptotes cross for $u = u_{c,b} = (12/\sqrt{2})^{1/3} \approx 2.04$.
\end{itemize}
This, for a given detector, takes place for a value of the momentum $p_u$ for which
\begin{equation}
u_c =
N\sqrt{\gfrac{l}{\sigma}\gfrac{p_0}{p}\sqrt{\gfrac{\Delta x}{X_0}}}
,
\end{equation}
that is, 
\begin{equation} \label{eq:pu}
p_u = p_0 \sqrt{\gfrac{\Delta x}{X_0}} \gfrac{N^2 l}{\sigma u_c^2},
\end{equation}
from which
\begin{equation} \label{eq:pu:param}
 u = u_c \sqrt{\gfrac{p_c}{p}} ,
 \ \ \
 p = p_u \left( \gfrac{u_c}{u} \right)^2.
\end{equation}

In short, a homogeneous detector is a thick detector, $u > u_c$, at low
momentum, $p < p_u$ and a thin detector at higher momentum.
In Table \ref{tab:detecteurs}, we use $u_c = 2.5$ to compute the value of $p_u$.

In the same way, a detector is a homogeneous detector, $x < x_c$ at high momentum,
$p > p_x$, with
\begin{equation} \label{eq:px}
p_x = p_0 \sqrt{\gfrac{\Delta x}{X_0}} \gfrac{l}{\sigma x_c^2}.
\end{equation}

and $x_c = 0.2$ (Fig. \ref{fig:V_x}).
And similarly:
\begin{equation} \label{eq:px:param}
 x = x_c \sqrt{\gfrac{p_c}{p}} ,
 \ \ \
 p = p_x \left( \gfrac{x_c}{x} \right)^2.
\end{equation}

\begin{table} \centering
 \caption{Parameters of two trackers considered in the text.
 \label{tab:detecteurs}}

~
 
\begin{tabular}{|c|c|c|c|c|c|c|c|} \hline
           & gas argon   & liquid argon     &  silicon  &  & \\ 
           &  TPC &  TPC   &  detector  &  & \\ \hline
  $X_0$    & 2351. &  14.0 &   9.4  &    $\centi\meter$ & \\
  $l$      &   0.1 &  0.3  &   1.0 &   $\centi\meter$ & \\
  $\Delta x$ &   $l$ &   $l$ &   0.0500 &  $\centi\meter$ & \\
  $\sigma$ &  0.1 &  0.1  &    0.0070 &  $\centi\meter$ & \\
  $L$      &  30.  & 1000.  &       &  $\centi\meter$ & \\
  $N$      &  300  & 3\,333  &   56  &  & \\
  $p_{1}$ & 0.112 &  1.739 &   0.239 &  $\mega\electronvolt/c$ & eq. (\ref{eq:p_1}) \\
  $p_u$    & 1277. & 10\,614\,042. &   71\,098.  &  $\mega\electronvolt/c$ & eq. (\ref{eq:pu}) \\
  $p_x$    &   2.2 &   149. &   3\,542. &  $\mega\electronvolt/c$ & eq. (\ref{eq:px}) \\
  $p_s$    & 0.024 &   5.4 &    16.6 &  $\mega\electronvolt/c$ & eq. (\ref{eq:ps}) \\
  $p_\ell$ &  352. & 2\,931\,742.  &    19\,638.   &  $\mega\electronvolt/c$ & eq. (\ref{eq:pell}) \\
\hline
\end{tabular}
\end{table}

\begin{itemize}
\item {\bf Argon gas TPC}. We see that $p_u > p > p_x$ for most of the
 [1\,MeV - 1\,GeV] momentum range that is the primary target of the
 high-performance $\gamma$-ray telescopes mentioned above: the
 telescope is both a homogeneous and a thick detector.
 
\item {\bf Silicon detector}. Here the telescope is a segmented and
 a thick detector for most of the momentum range.
\end{itemize}

Note that the equality $p_u = p_x$ holds for $N = u_c / x_c$,
 that is, $N = 2.5 / 0.2 = 12.5$, so for most conceivable detectors,
 $p_u > p_x$, that is, 
\begin{itemize}
\item if $p > p_u$ then $p > p_x$,
 if a detector is thin for a given track, then it's also homogeneous;
\item if $p < p_x$ then $p < p_u$,
 if a detector is segmented for a given track, then it's also thick.
\end{itemize}

\section{Kalman filter}
\label{sec:kalman}

A Kalman filter is an estimator of the state of a
linear dynamic system perturbed by Gaussian white noise using
measurements that are linear functions of the system state and
corrupted by additive Gaussian white noise \cite{Grewal:2001}.
The paraxial propagation of a high-momentum particle inside a detector
is affected by angular deflections due to scattering on the charged
particles (electrons, nuclei) present in the detector matter.
Deflections undergone at different locations on the track are
uncorrelated, ``white process noise'', and are approximated to have a 
Gaussian distribution under the multiple-scattering approximation.
Transverse 
position measurements are performed at several locations along the
track. They are affected by an uncertainty that does not correlate
from layer to layer, ``white measurement noise'', and that most often
can be approximated by a Gaussian distribution.
Angular deflections and measurement uncertainty are not correlated.
When the system is non-linear, such as for the propagation in a
magnetic field, it is linearized locally, ``extended Kalman filter''
(\cite{Fujii} and references therein).
In the case of most particle detectors, the geometric, the multiple
scattering and the measurement properties of the detector are uniform (at
least piecewise) so the dynamics of the particle is described by a
time-invariant system.

Since the founding work by Frühwirth \cite{Fruhwirth:1987fm}, KF
tracking has been used largely in high-energy physics.
We present here a short description of the elements that are
used in the next section, in a Bayesian formulation.
Denoting $\{z^0_n\}$ and $\{z^{m}_n\}$ the true and the measured
positions of a particle at layer $n$,
respectively, and $x_n$ the corresponding state vector,
$\hat x_n=\E(x_n|z_0^m,\cdots,z_n^m)$ is the estimator of $x_n$
conditioned to $\{z_0^m,\cdots,z_n^m\}$
and $x_n^{n-1}=\E(x_n|z_0^m,\cdots,z_{n-1}^m)$ is the prediction of
$x_n$ given $\{z_0^m,\cdots,z_{n-1}^m\}$.
$x_n$ is obtained from $x_{n-1}$ 
\begin{equation}\label{eq:propagation:xn}
 x_n=D\cdot x_{n-1}+D\cdot \begin{bmatrix}0\\u_n\end{bmatrix}
 ;
\end{equation}

$u_n$ is the Gaussian-distributed deflection angle with variance $sl$.
The covariance matrix of the state vectors is
$P_n=\E((\hat x_n-x_n)(\hat x_n-x_n)^T|z_0,\cdots,z_n^m)$.
The optimal estimator of $x_n$ is obtained from $\hat x_{n-1}$ and
from the measurements $\{z_0^m,\cdots,z_{n-1}^m\}$,
\begin{eqnarray}\label{eq:optimal:xn}
 x_n^{n-1}&=&D\hat x_{n-1}
 ,\\
 P^{n-1}_n&=&D(P_{n-1}+B)D^T
 ,\label{eq:Pn}
\end{eqnarray}
with
\begin{equation}\label{eq:zn}
z_m^n=Hx_n+v_n ,
\end{equation}
where $H=\begin{bmatrix}0&1\end{bmatrix}$ is the
 measurement matrix and $v_n$ is the measurement
 uncertainty which is Gaussian-distributed with variance $\sigma^2=\gfrac{1}{\imath l}$.
The innovations are the difference between measurement and prediction, 
\begin{eqnarray}\label{eq:nun}
\nu_n&=&z_n^m-x_n^{n-1}
\end{eqnarray}
and their variance is 
\begin{eqnarray}
S_n=\cov(\nu_n)&=&\sigma^2+HP_n^{n-1}H^T . \label{eq:Sn}
\end{eqnarray}

The gain matrix of the filter is 
\begin{equation}\label{eq:Kn}
 K_n=P_n^{n-1}H^TS_n^{-1}
 .
\end{equation}

For the optimal value of the gain that minimizes the variance of the
innovations, we obtain \cite{Grewal:2001}
\begin{eqnarray}\label{eq:Kalman:optimal:xn}
\hat x_n&=&x_n^{n-1}+K_n\nu_n , \\
P_n&=&P_n^{n-1}-K_nS_nK_n^T.
\end{eqnarray}

Noting $Z^n=z_0^m,\cdots,z_n^m$ the set of measurements up to layer $n$
and $p$ the probability density, 
\begin{eqnarray}
 p(Z^n)&=&p(z_n^m,Z^{n-1}) \nonumber
 \\
 &=&p(z_n^m|Z^{n-1})p(Z^{n-1})\ \ \ \ \ \text{(Bayes)} \nonumber
 \\
 &=&\prod_{i=0}^np(z_i^m|Z^{i-1})\ \ \ \ \ \text{(recurrence)}
 . \label{eq:pn}
\end{eqnarray}

As $z_i^m|Z^{i-1}$ is Gaussian distributed ${\mathcal N}(z_{i}, S_i)$:
\begin{eqnarray}
 p(z_i^m|Z^{i-1}) & =
 &\gfrac{1}{\sqrt{2\pi S_i}}\exp\left[-\frac{\nu_i^2}{2S_i}\right] . \label{eq:p:normal}
\end{eqnarray}

\begin{figure}
\includegraphics[width=0.49\linewidth]{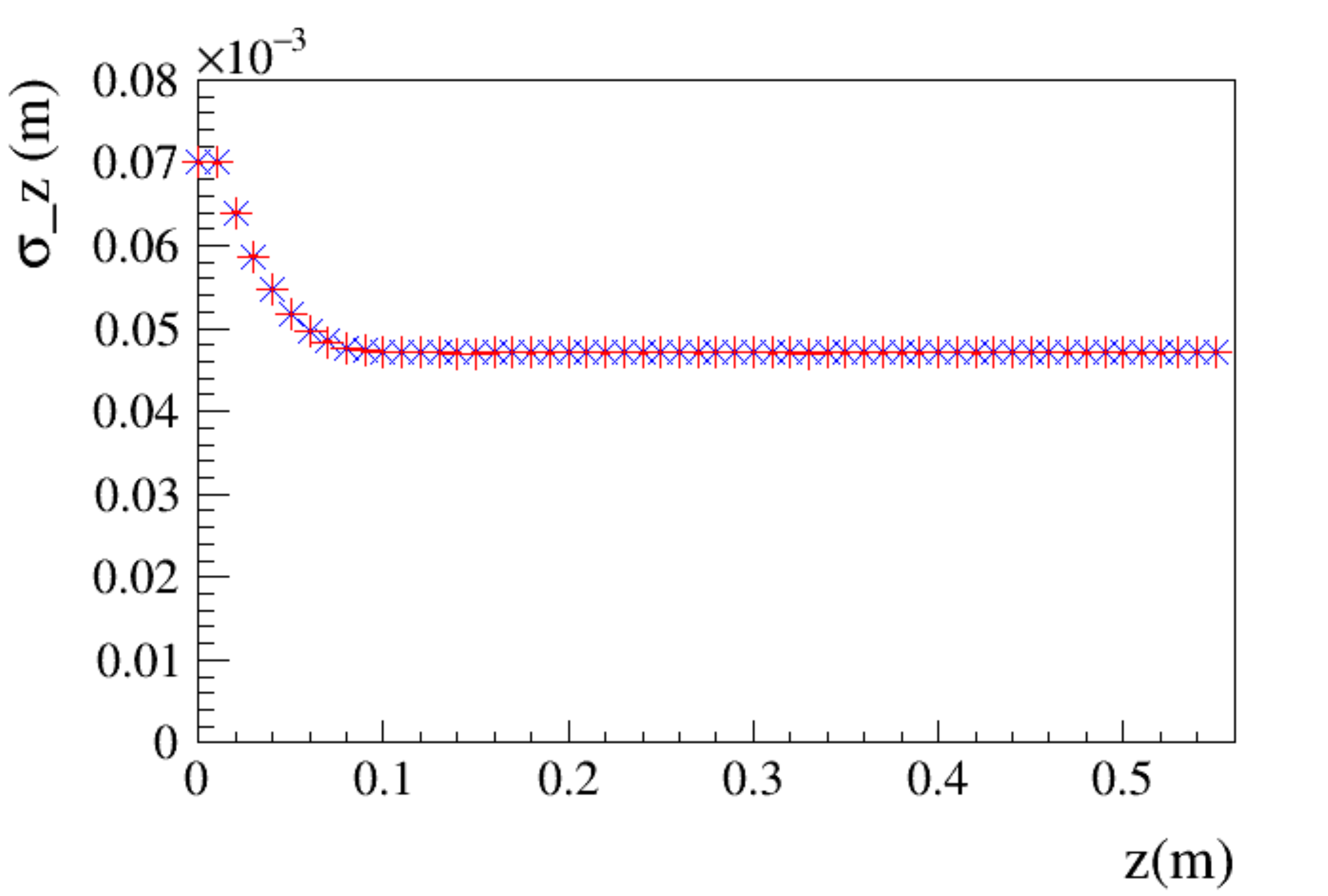}
\includegraphics[width=0.49\linewidth]{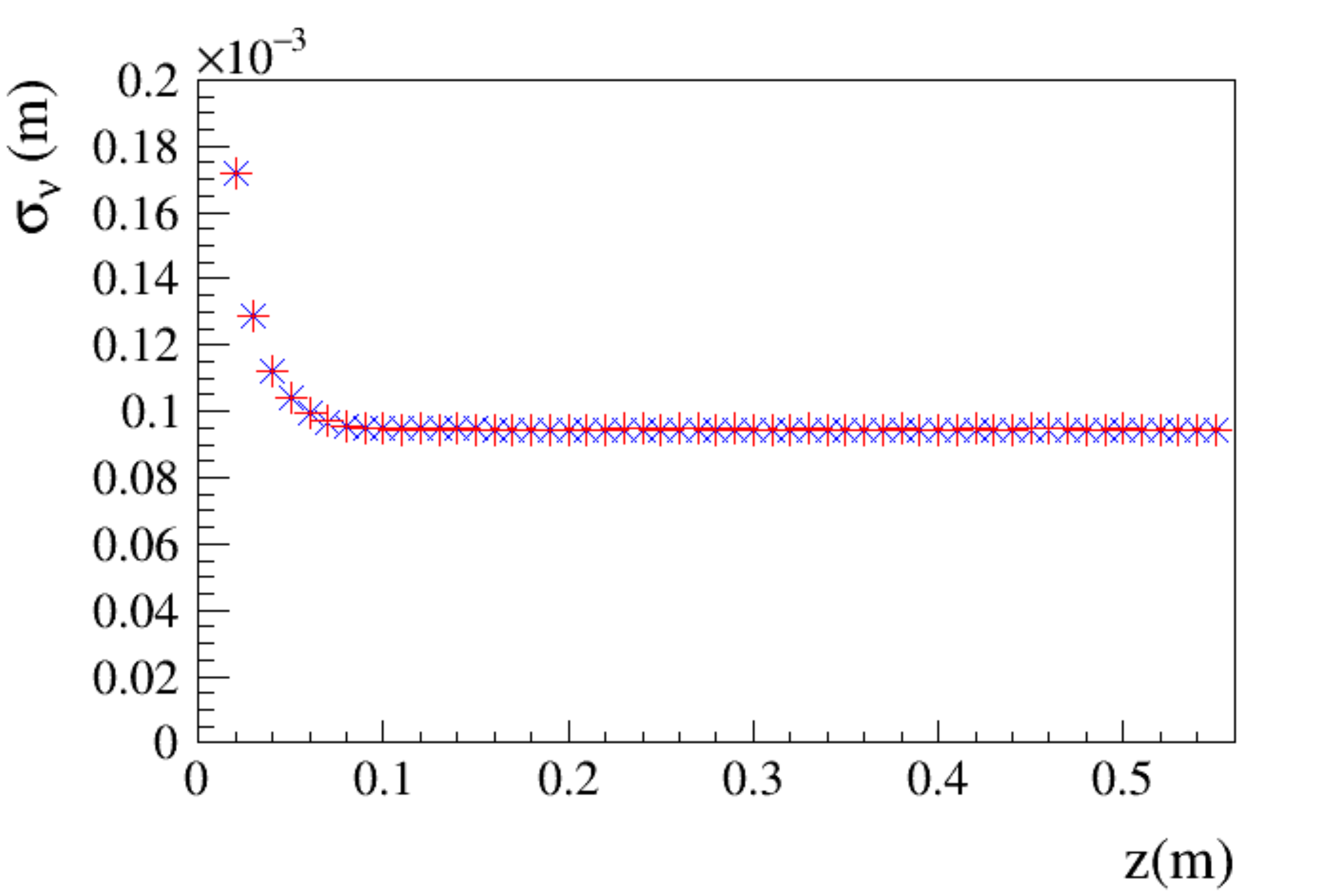}
\caption{\label{g:kal_val}
Kalman filter validation: RMS (plusses) of the position residues (left) and
 of the
innovation residues (right) as a function of the longitudinal position in the
tracker, for a sample of $10^6$ $50\,\mega\electronvolt/c$ tracks in a
silicon detector, compared to the RMS (crosses) computed from their
variances, $P_n$ and $S_n$, respectively. }
\end{figure}

We have implemented such a KF tracking software. Figure
\ref{g:kal_val} shows a couple of sanity-check validation plots, the
RMS of the position residues (left) and of the innovation residues (right) as a
function of the longitudinal position in the tracker, for a sample of
$10^6$ $50\,\mega\electronvolt/c$ tracks in a silicon detector,
compared to the RMS computed from their variance, $P_n$ and $S_n$,
respectively.

\section{Momentum Measurement}
\label{sec:momentum:measurement}

A Kalman filter is the optimal linear estimator of the state vector of
a dynamical system at the condition that the model be an accurate
description of the dynamics of the system and that the process and
measurement noise covariance matrices be known, that is 
here, that the track momentum be known.
The estimation of the noise covariance matrices of a dynamic system
was pioneered by Mehra \cite{Mehra:1970,Mehra:1972}
who studied and compared several methods:
\begin{itemize}
\item A Bayesian method that is the root of that we use in this work.
 
\item Maximum likelihood methods, if necessary of both the state vector of the system and the noise matrices at the same time.
\end{itemize}
Bayesian and maximum likelihood methods were deemed to be too CPU consuming for the time.
 
\begin{itemize}
\item Covariance-matching techniques, making the innovation residuals
 consistent with their theoretical covariances; these methods were
 shown later to give biased estimates of the covariance matrices.

\item Correlation methods, in particular based on the observation that
 when the KF gain $K$ is optimal, the innovations of the filtering
 process are white and Gaussian.
\end{itemize}
Mehra showed that the optimal gain $K$ can be determined uniquely,
after which many efforts and publications have then been spent in
determining the convergence of these methods and to which values of
the process and measurement noise matrices they were, eventually,
converging.
In 2006, Odelson {\it et al.} \cite{Odelson:2006} re-examined Mehra's work,
showed that the definite positiveness of the matrices was not
assured;
based on the fact that the autocorrelation of the innovation
sequence is linearly dependent on the noise covariances
\cite{Belanger:1974},
 they developed an autocovariance least-square (ALS) method that provides
unbiased estimates of the noise matrices and that includes a mechanism that
enforces definite positiveness \cite{Odelson:2006}.

Kalman filters have already been used for momentum measurement in
non-magnetic particle physics detectors in the past
(\cite{Stanco:1989mg,Antonello:2016niy}.
The trick is to augment the state vector with the parameter vector to
($x, y, \dd x/\dd z, \dd y/\dd z, 1/p$) so that the KF performs their
 estimation simultaneously.
However, this augmentation approach has been originally intended for
estimating parameters in deterministic part of the model and its
straightforward application for noise covariance matrices does not
result in appropriate estimates (\cite{Dunik:2009} and references
therein).

In the case of charged particle tracking in a magnetic-field-free
detector, the augmentation method was found to provide unbiased
results though \cite{Stanco:1989mg,Antonello:2016niy}, most likely as
the tracking part and the deflection part of the filter behave as two
separate filters, ``only'' linked by the joint uses of the track
momentum, one for the process noise matrix, the other as part of the
state vector.
Also it enables the optimal treatment of energy loss and therefore
it
provides an improvement of about a factor of two with respect to the
Molière method \cite{Antonello:2016niy}.
If would be interesting to examine to what extent they are efficient
or even whether they are optimal.
Note that in that scheme, the track has to be segmented to measure the track
angle on each segment
(in $\approx 19\,\centi\meter$ long segments that contain $\approx 57$
hits on average for\cite{Antonello:2016niy}, from which they obtain a
relative resolution of 16\,\% on a sample of $4\,\meter$ tracks
with momenta ranging from 0.5 to $4.5\,\giga\electronvolt/c$).
(See also \cite{Abratenko:2017nki}).

\subsection{Single track momentum measurement: Bayesian method}
\label{sub:sec:bayesian:method}

Following Matisko and Havlena \cite{Matisko:Havlena:2013} we 
 obtain \footnote{
We assume that the detector spatial resolution $\sigma$ is known
either from calibration on beam or from the analysis of high
momentum tracks.},
from the measurements, the most probable value of $s$, and
we extract from it an optimal estimator $\hat p$ of $p$.
For an event $A$, defining $p_n(A) \equiv p(A|Z^n)$, we have 
\begin{eqnarray}
 p_n(s)&=
&\frac{p_{n-1}(s,z_n^m)}{p_{n-1}(z_n^m)} \nonumber
 \\
 &=&\frac{p_{n-1}(s)p_{n-1}({z_n^m|s})}{p_{n-1}(z_n^m)}
 . \label{eq:pn:2}
\end{eqnarray}

We name $s$-filter a KF with a gain matrix computed with a given value
of $s$.
We remember (eq.~(\ref{eq:Sn})) that the $s$-filter innovation
probability density function (pdf) $\beta_n=p_{n-1}(z_n^m|s)$,
is a normal, $\beta_n=\mathcal N(\nu_n(s),0,S_n(s))$, where $\nu_n(s)$ and $S_n(s)$
are computed during the filtering process.
The $1/p_{n-1}(z_n^m)$ factor does not vary with $s$ and
is therefore neglected.
We obtain 
\begin{equation}\label{eq:pn:3}
 p_n(s)\propto \prod_i\beta_i .
\end{equation}

\begin{figure}[h]
\centering 
\includegraphics[width=0.475\textwidth]{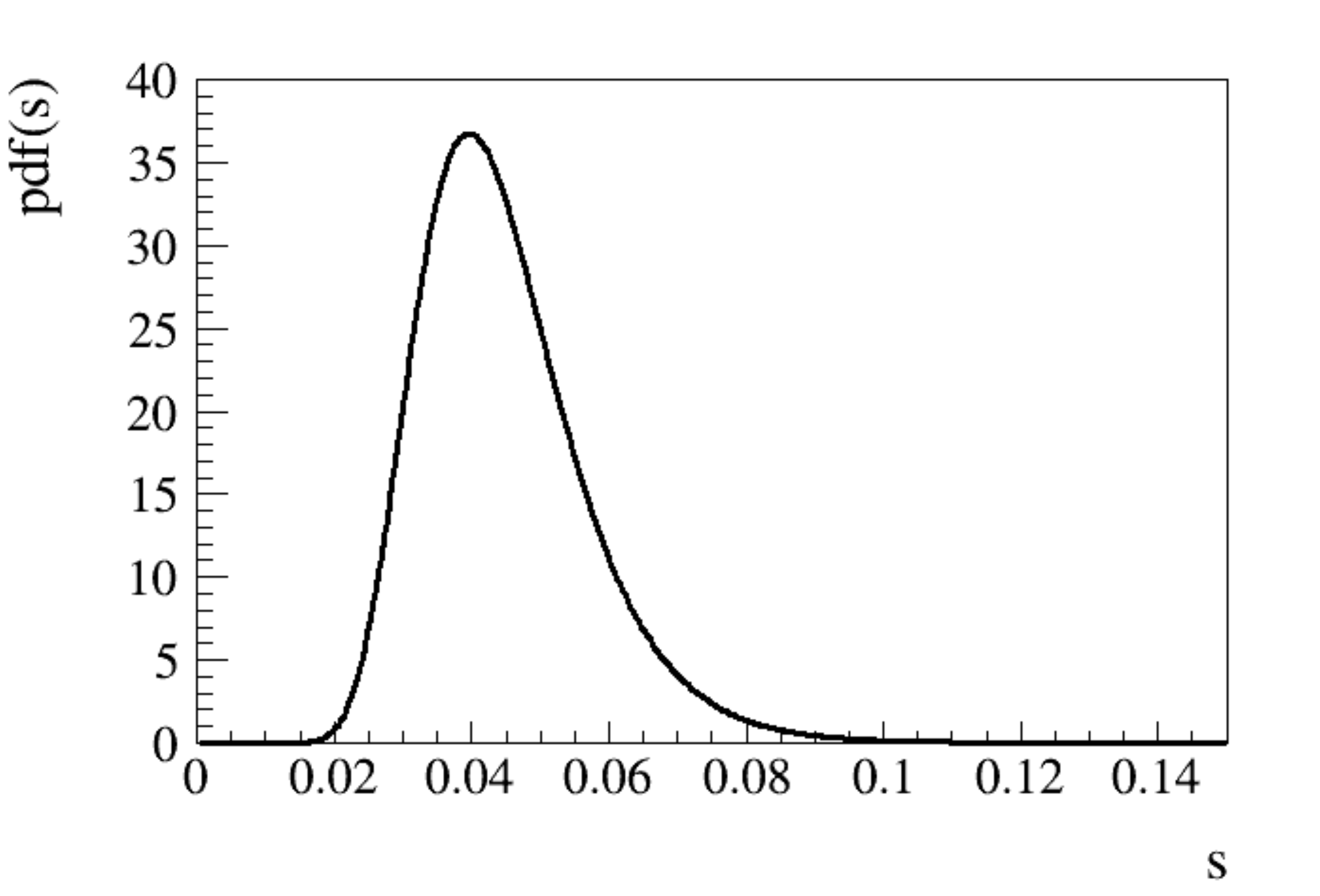}
\hfill
\includegraphics[width=0.475\textwidth]{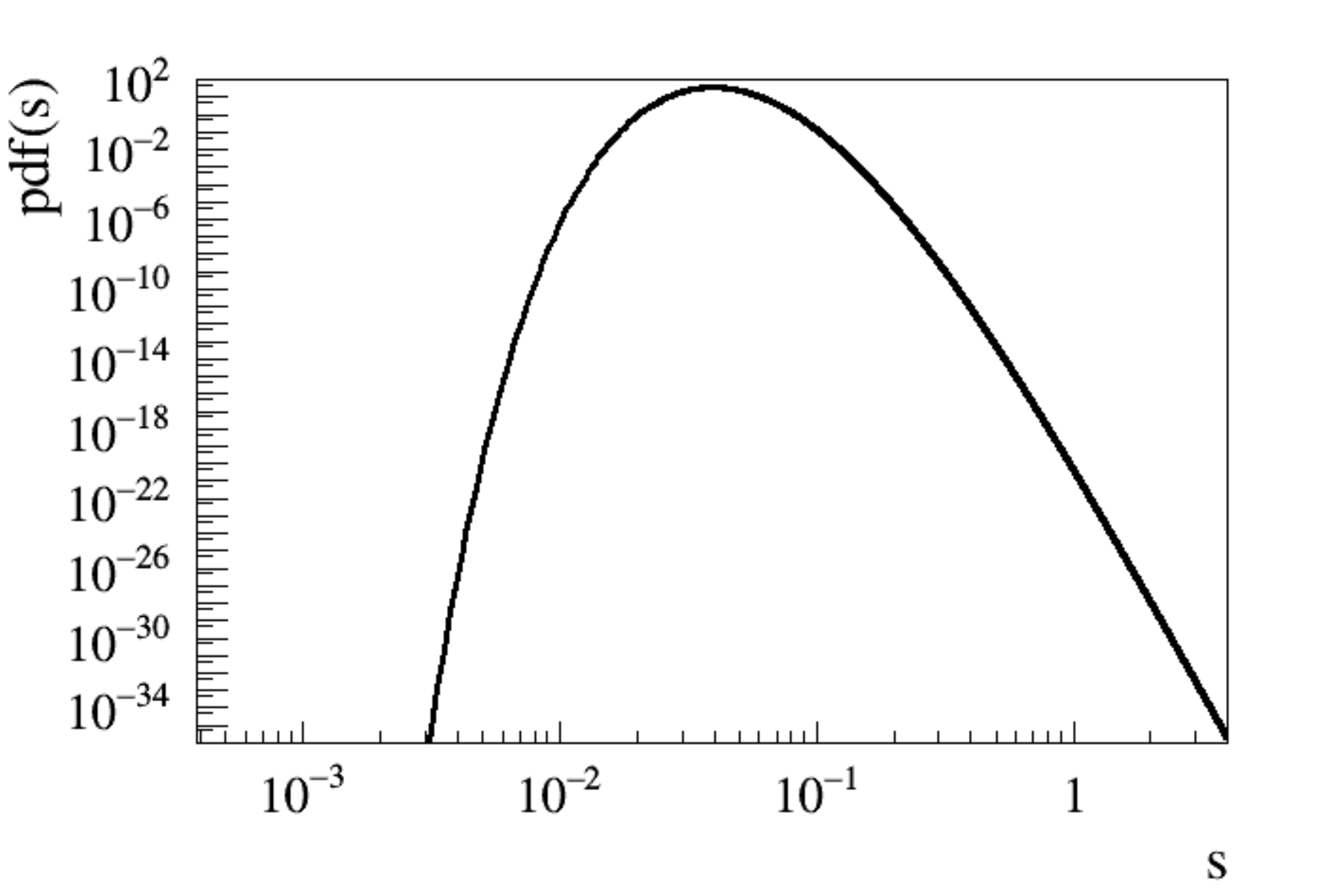}
\caption{\label{fig:p:s}$p(s)$ distribution for a $50\,\mega\electronvolt/c$ track in a silicon detector (eq. (\ref{eq:pn:3})).
 On that track, the momentum is measured to be equal to
 $49.9\,\mega\electronvolt/c$.
Linear (left) and logarithmic (right) scales. }
\end{figure}

The distribution of $p(s)$ for one simulated $50\,\mega\electronvolt/c$ track is shown in Fig. \ref{fig:p:s}.
The track momentum is then obtained from the value of $s$ that
maximizes $p_n(s)$: 
\begin{equation}
 p = p_0 \sqrt{\gfrac{\Delta x}{l X_0 s}}
 .
\end{equation}

\begin{figure}[h]
\includegraphics[width=0.45\linewidth]{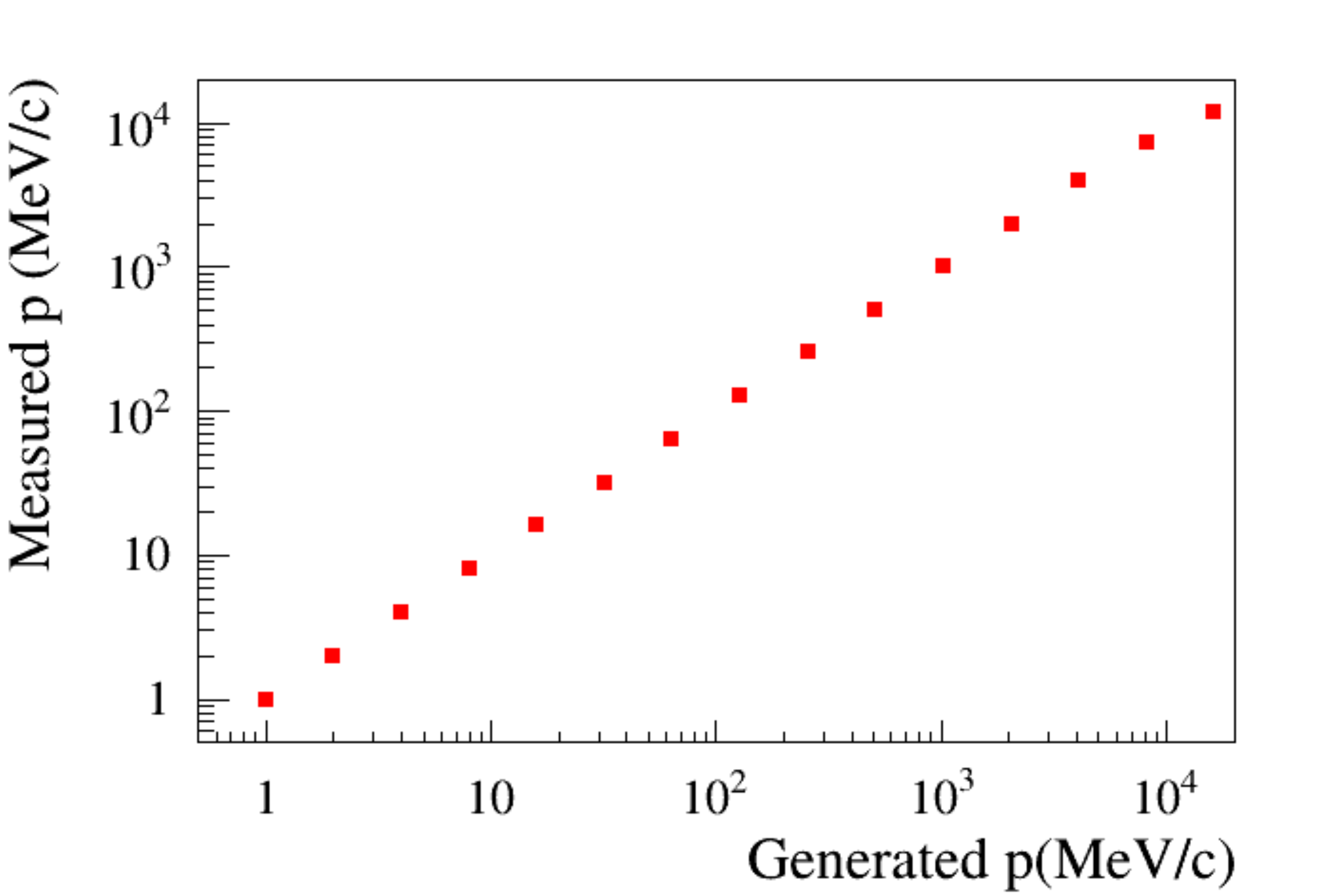}
\includegraphics[width=0.45\linewidth]{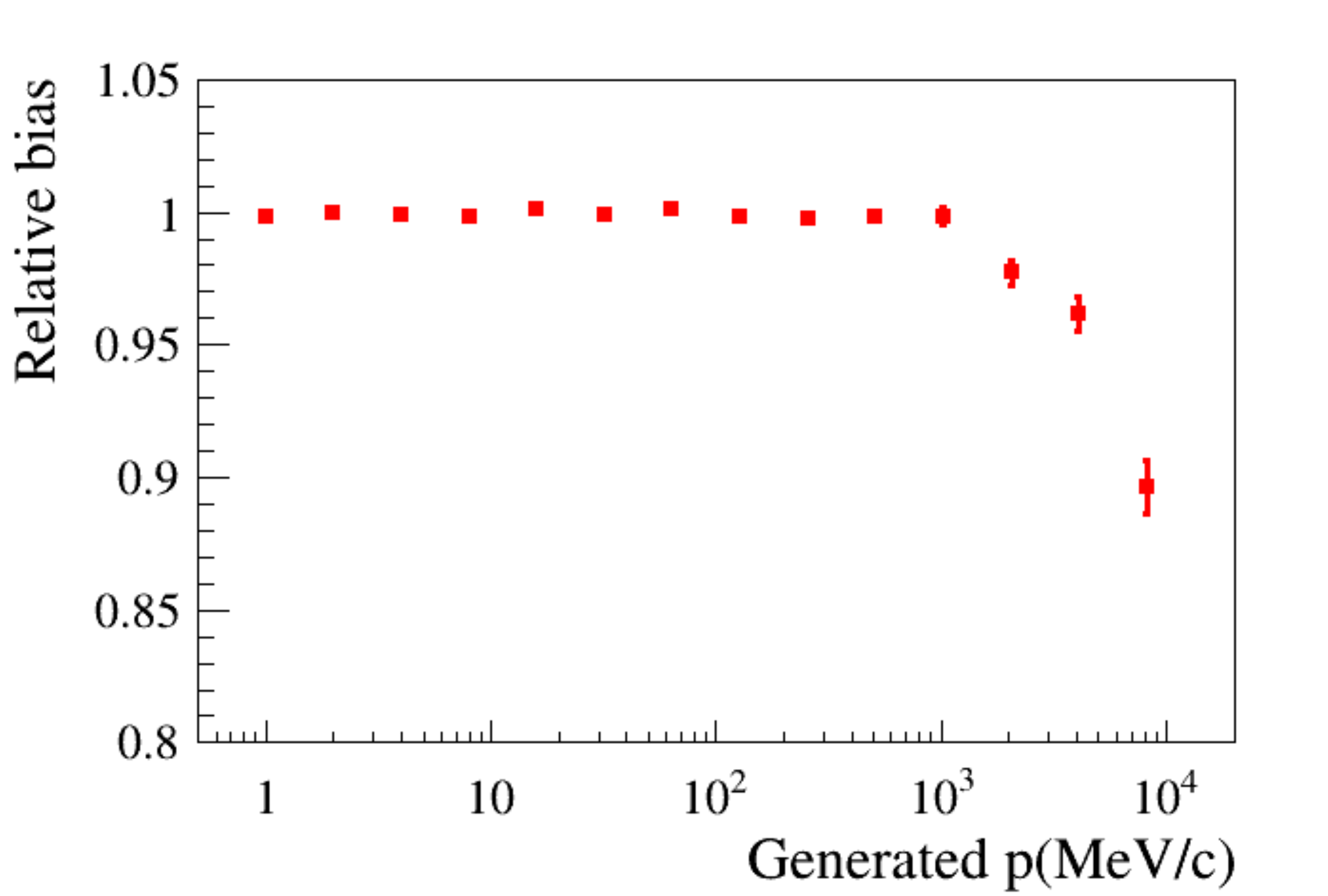}
 \put (-280,80) {(a)}
 \put (-30,80) {(b)}

\includegraphics[width=0.45\linewidth]{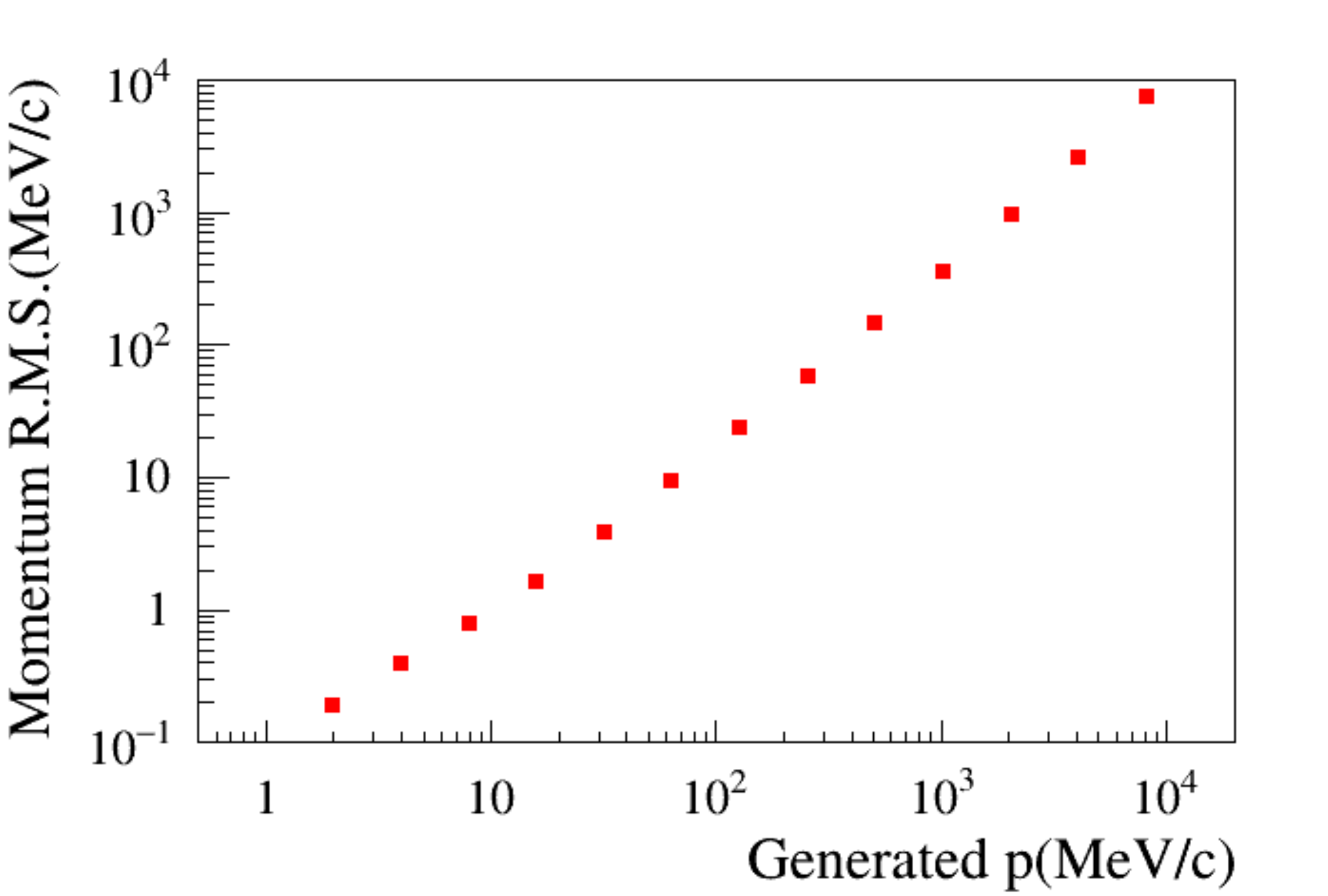}
\includegraphics[width=0.45\linewidth]{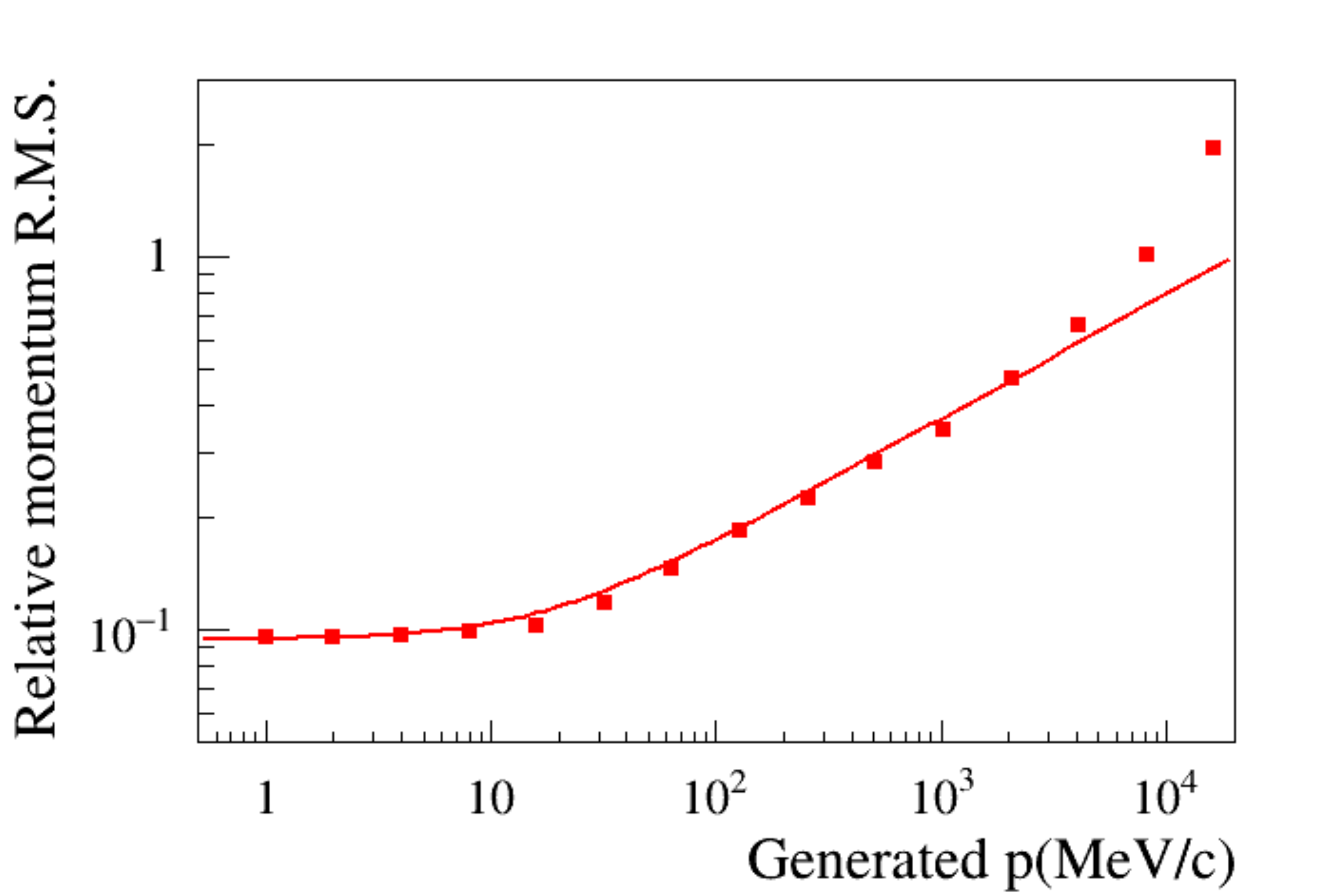}
 \put (-280,80) {(c)}
 \put (-50,80) {(d)}

\caption{\label{fig:silicium}
 Performance of the momentum measurement for the silicon detector: 
 Variation as a function of the true (generated) particle momentum of
 (a) the average measured momentum;
 (b) the average measured normalized to the generated momentum;
 (c) R.M.S of the measured momenta;
 (d) the relative R.M.S of the measured momenta.
The curve is from eq. (\ref{eq:sigma:sur:p:parametrique}).
}
\end{figure}

\begin{figure}[h]
\includegraphics[width=0.45\linewidth]{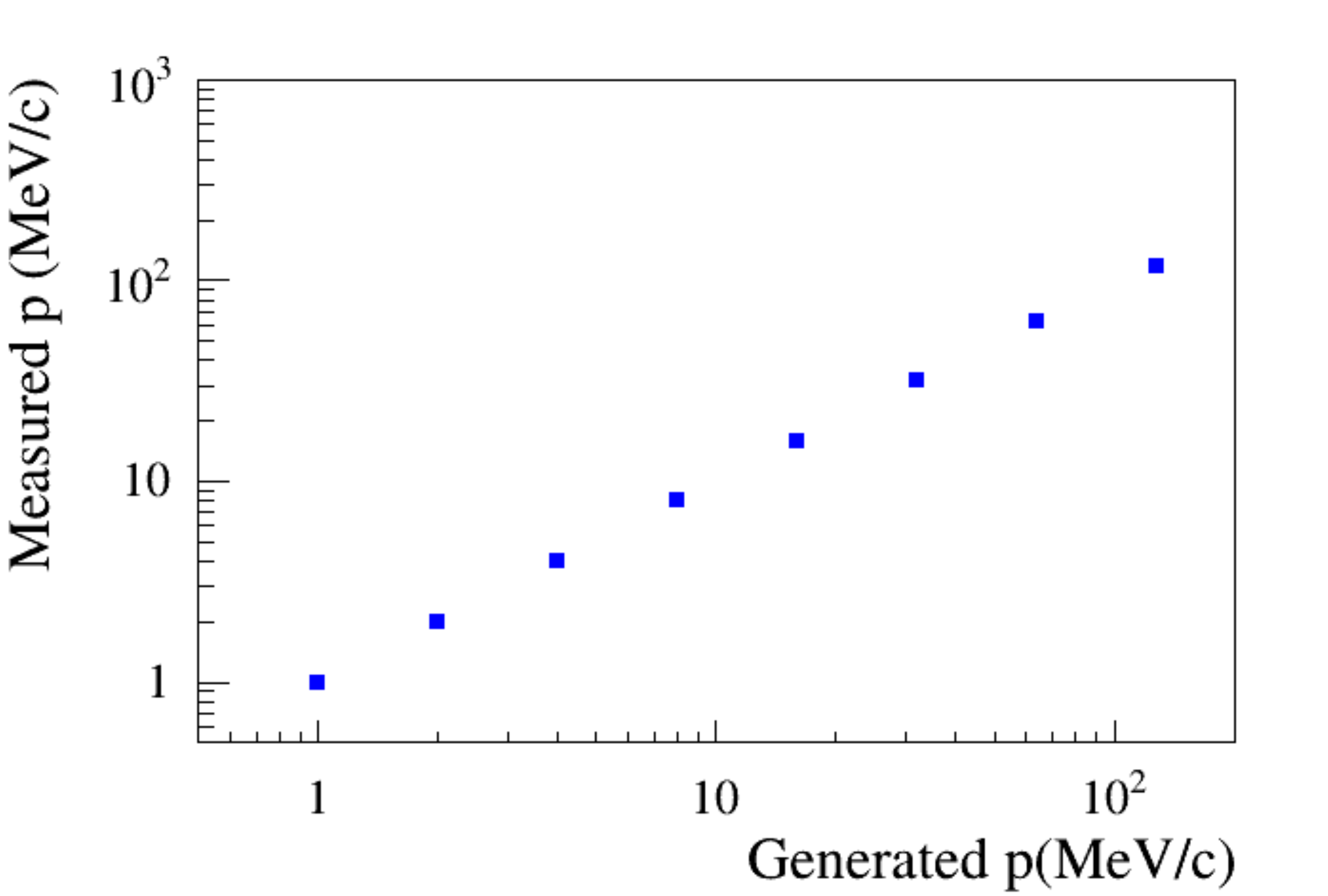}
\includegraphics[width=0.45\linewidth]{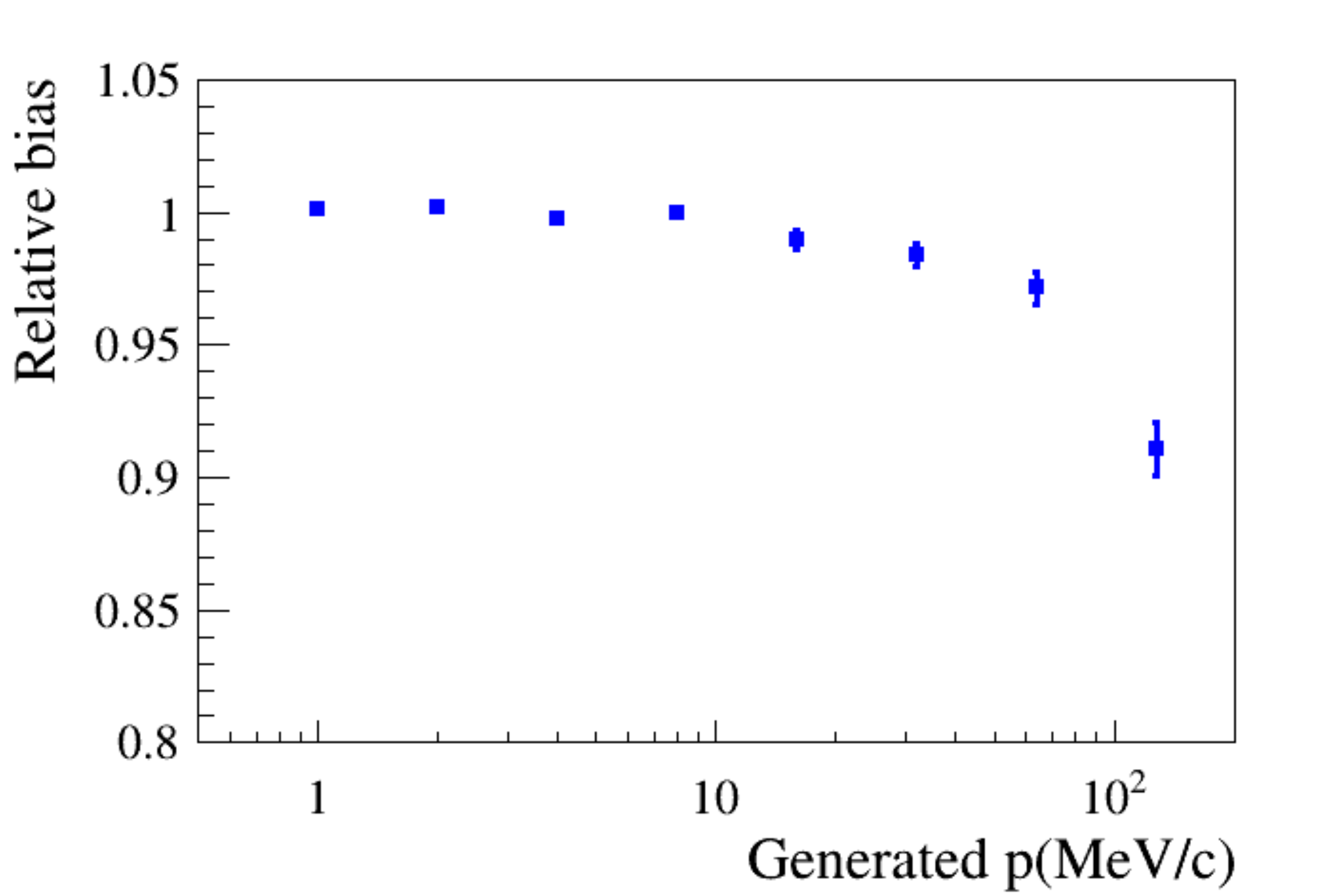}
 \put (-280,80) {(a)}
 \put (-30,80) {(b)}

\includegraphics[width=0.45\linewidth]{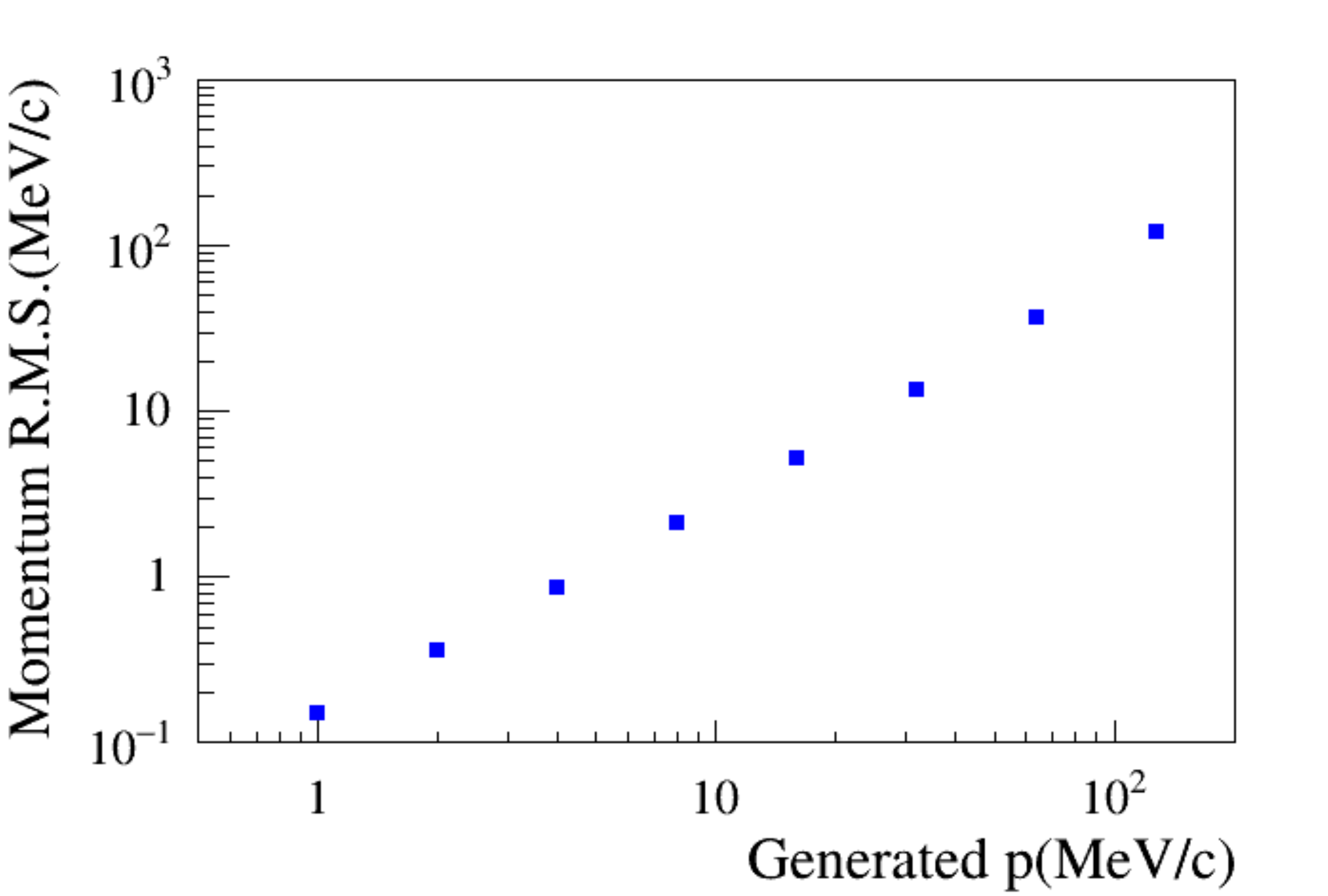}
\includegraphics[width=0.45\linewidth]{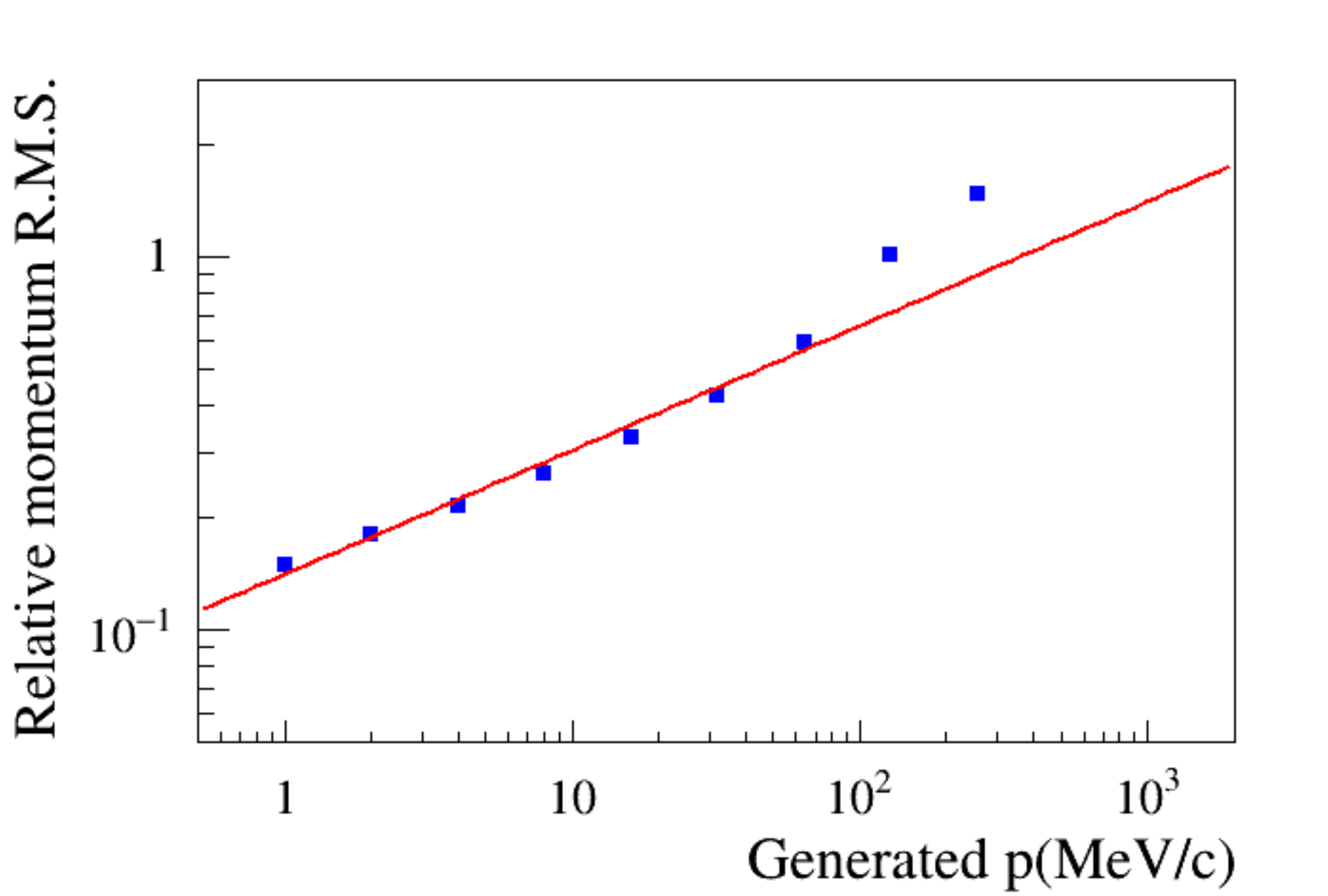}
 \put (-280,80) {(c)}
 \put (-30,50) {(d)}

\caption{\label{fig:argon}
 Performance of the momentum measurement for the argon gas detector: 
 Variation as a function of the true (generated) particle momentum of
 (a) the average measured momentum;
 (b) the average measured normalized to the generated momentum;
 (c) R.M.S of the measured momenta;
 (d) the relative R.M.S of the measured momenta.
The curve is from eq. (\ref{eq:sigma:sur:p:parametrique}).
}
\end{figure}

\begin{figure}[h]
\includegraphics[width=0.45\linewidth]{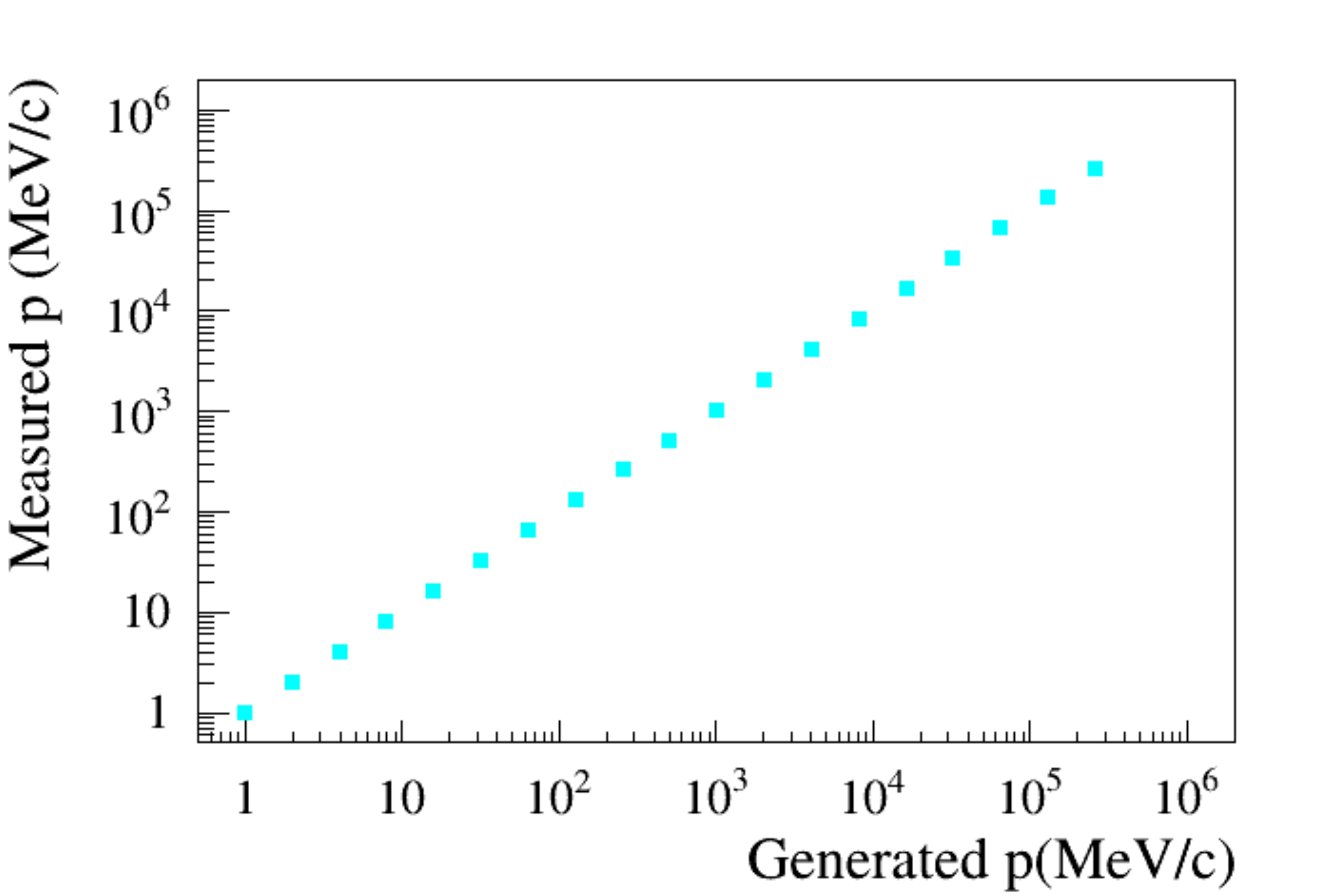}
\includegraphics[width=0.45\linewidth]{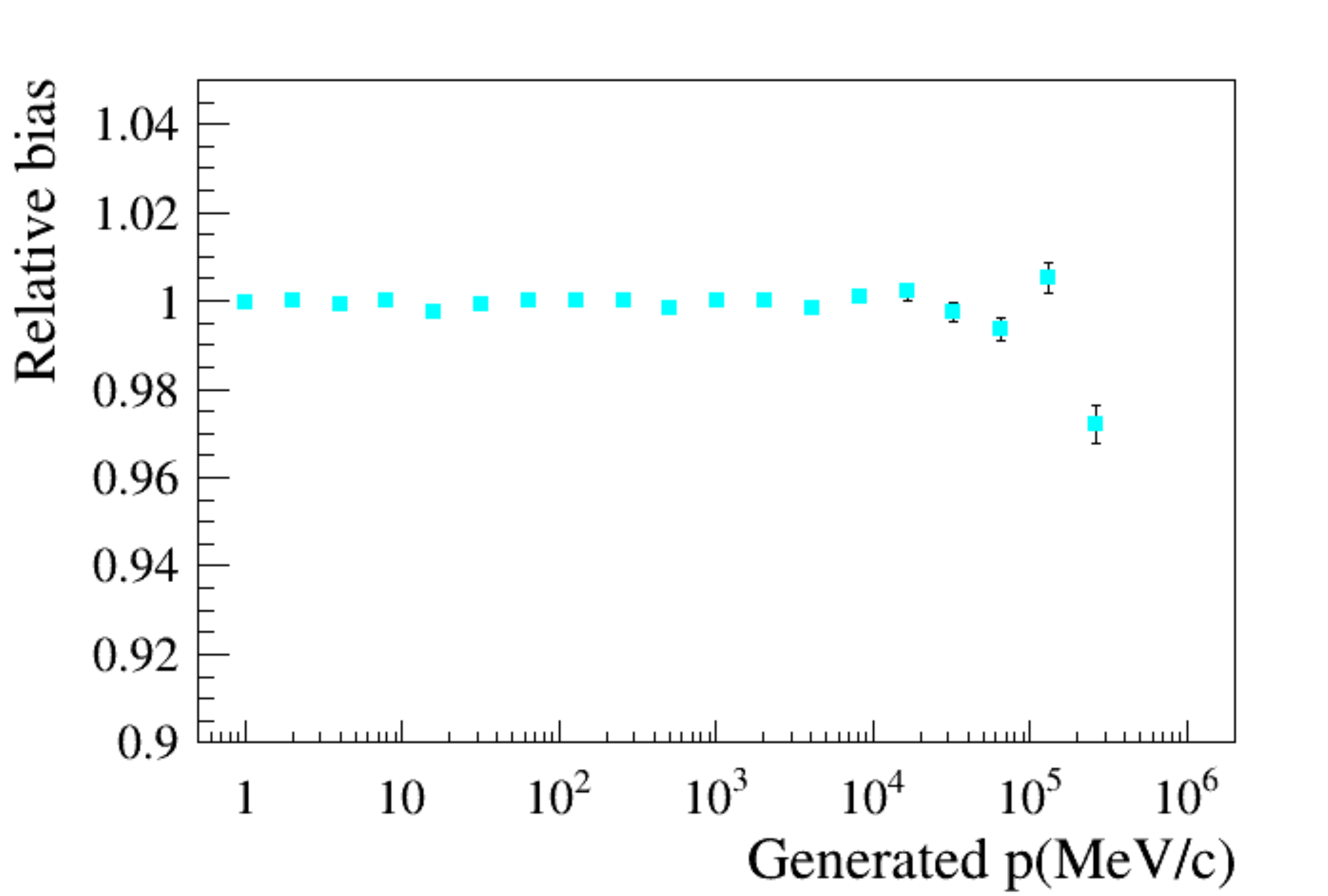}
 \put (-280,80) {(a)}
 \put (-30,80) {(b)}

\includegraphics[width=0.45\linewidth]{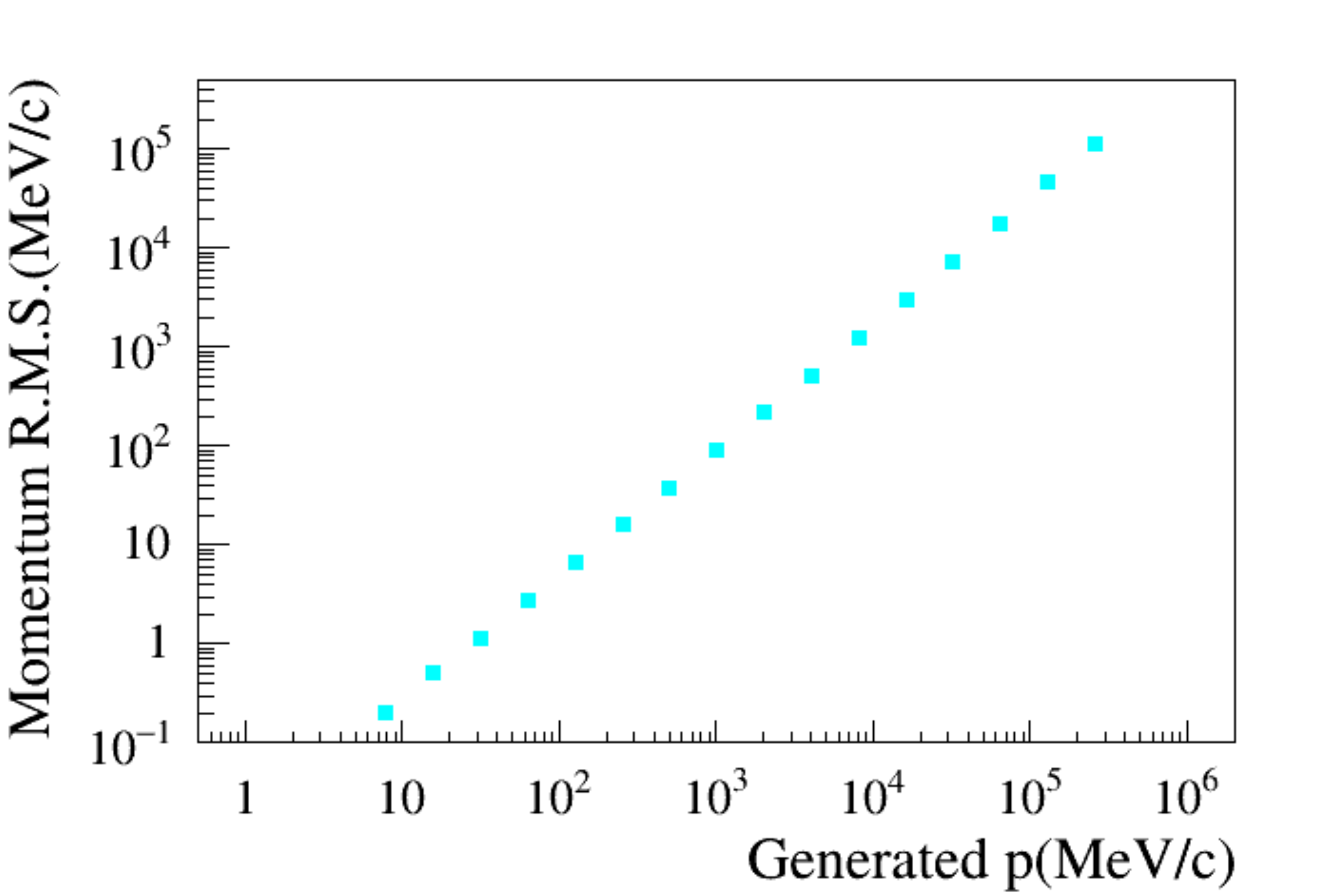}
\includegraphics[width=0.45\linewidth]{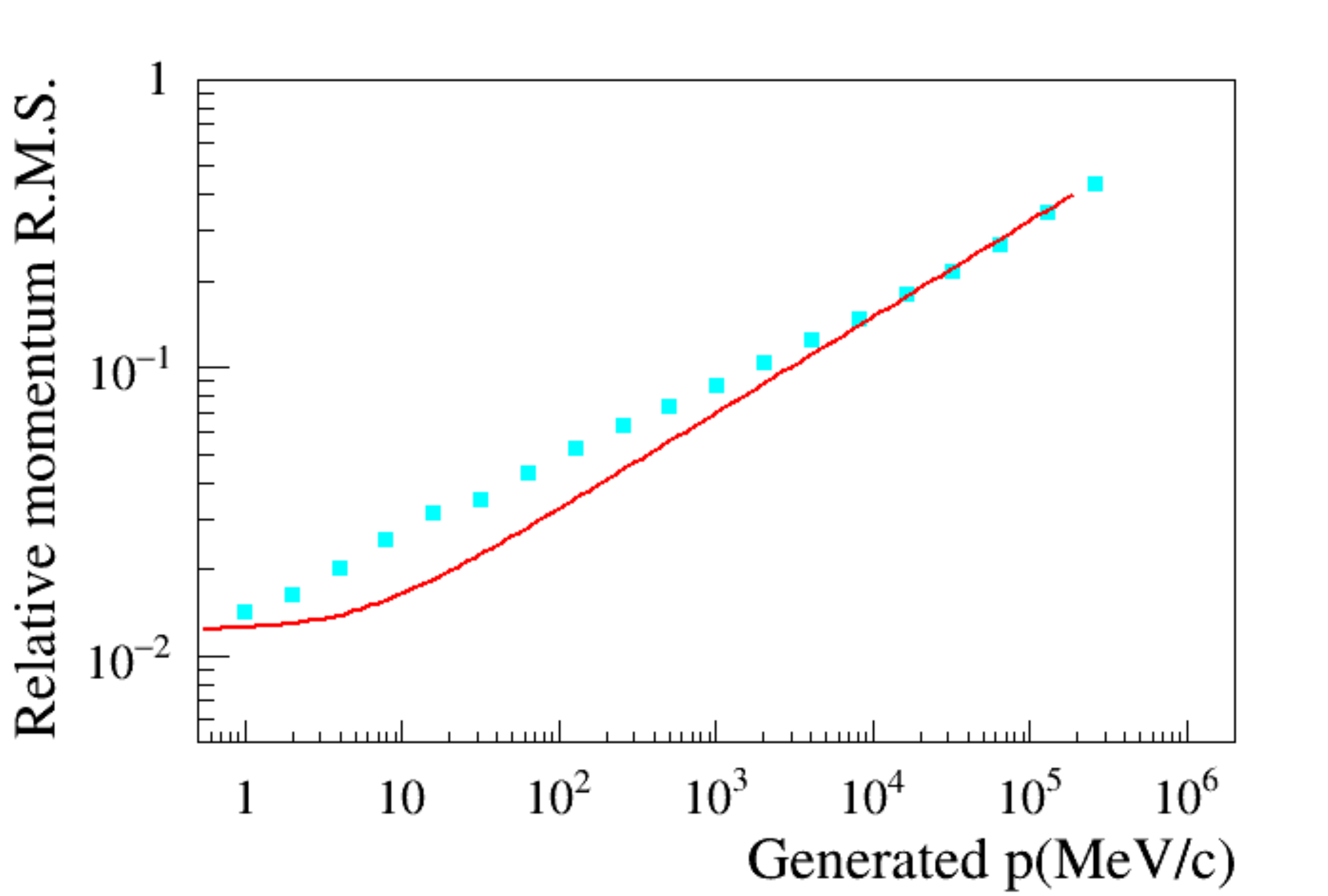}
 \put (-280,80) {(c)}
 \put (-30,50) {(d)}

\caption{\label{fig:argon:liq}
 Performance of the momentum measurement for the liquid argon detector: 
 Variation as a function of the true (generated) particle momentum of
 (a) the average measured momentum;
 (b) the average measured normalized to the generated momentum;
 (c) R.M.S of the measured momenta;
 (d) the relative R.M.S of the measured momenta.
The curve is from eq. (\ref{eq:sigma:sur:p:parametrique}).
The discrepancy between the data and the curve for this large-$n$
detector, at low momentum that is at very low $\sigma_p/p$, needs
further investigation.
}
\end{figure}

From the full width half maximum (FWHM) of $p(s)$ we calculate
$\RMS_s/s$ and $\RMS_p/p = (\RMS_s/s)/2$.
The average value of $\RMS_p/p$ is found to be much smaller than the
relative RMS $\sigma_p/p$ of the momentum measurements performed on a
sample of tracks and shown in Figs. \ref{fig:silicium} and
\ref{fig:argon}.
Our interpretation is that the range of $s$ values compatible within
uncertainties with the deflection sample of a given track and the
fluctuation of the deflection sample from track to track are separate
quantities.
In addition, we observe that the single-track $p(s)$ width and the
measured momentum for that track are weakly correlated, so the former
would add little information to the measurement of the latter.

We were not able to obtain an analytical expression for the relative
precision of the momentum measurement.
Instead we performed a parametric variation study for a silicon
detector with
\begin{itemize}
\item
 $l = 0.5, 1.0, 2.0\,\centi\meter$,
\item
 $X_0 = 4.685, 9.37, 18.74\,\centi\meter$,
\item
 $N = 23, 46, 56, 92$, 
\item
 $\sigma^2 = 2.5, 5.0, 10.0 \times 10^{-5}\,\centi\meter^2$,
\item
 $p = 1 \cdots 2048\,\mega\electronvolt/c$
\end{itemize}
and with $\Delta x = 500\,\micro\meter$.
A good representation of these data is obtained with the following expression: 
\begin{equation}\label{eq:sigma:sur:p:parametrique}
 \gfrac{\sigma_p}{p} \approx \gfrac{1}{\sqrt{2 N}}
 \sqrt[4]{
1 + 256
 \left( \gfrac{p}{p_0} \right)^{4/3}
 \left( \gfrac{\sigma^2 X_0}{N \Delta x ~ l^2} \right)^{2/3}
} ,
\end{equation}
\begin{itemize}
\item
from which we obtain the obvious low-momentum asymptote 
\begin{equation}\label{eq:sigma:sur:p:parametrique:LE}
 \gfrac{\sigma_p}{p} \approx \gfrac{1}{\sqrt{2 N}}
\end{equation}
\item
 and the high-momentum asymptote 
\begin{equation}\label{eq:sigma:sur:p:parametrique:HE}
 \gfrac{\sigma_p}{p} \approx \sqrt{\gfrac{8}{N}}
 \left( \gfrac{p}{p_0} \right)^{1/3}
 \left( \gfrac{\sigma^2 X_0}{N \Delta x ~ l^2} \right)^{1/6}
 .
\end{equation}
\end{itemize}
Of particular interest is the momentum, $p_s$, above which 
$\sigma_p/p$ starts to depart from the low momentum asymptote,
\begin{equation}\label{eq:ps}
p_s = p_0 
 \gfrac{1}{64}
 \left( \gfrac{N\Delta x ~ l^2}{\sigma^2 X_0} \right)^{1/2}
 .
\end{equation}
We define also 
 the momentum, $p_l$, above which $\sigma_p/p$ is larger than
unity, which means that the measurement becomes meaningless:
\begin{equation}\label{eq:pell}
 p_\ell = p_0
 \left( \gfrac{N}{8}\right)^{3/2}
 \left( \gfrac{N \Delta x ~ l^2} {\sigma^2 X_0}\right)^{1/2}.
\end{equation}
The only thing that can be said then is that that track is a straight
track within uncertainties, that is, with inverse momentum $1/p$
compatible with zero.
These two momenta are characteristics of the ability to measure
track momenta with a given detector and are related to each other, 
\begin{equation}\label{eq:pell:bis}
 p_\ell = p_s
 \left( {2 N}\right)^{3/2}
 .
\end{equation}

Finally, we obtain a simpler expression of the relative momentum resolution, 
\begin{equation}\label{eq:sigma:sur:p:param:simple}
 \gfrac{\sigma_p}{p} \approx
 \gfrac{1}{\sqrt{2 N}} \sqrt[4]{1 + \left( \gfrac{p}{p_s} \right)^{4/3}}
\end{equation}

The target relative precision of the DUNE project of 18\,\% is within
reach for $10\,\meter$ tracks up to a momentum of $17.1\,\giga\electronvolt/c$
 with detector parameter values from Table \ref{tab:detecteurs}
(Fig. \ref{fig:argon:liq}).

\subsection{Comparison with the cell-optimization result}

For the continuous detector, $\Delta x = l$, eq. (\ref{eq:sigma:sur:p:parametrique:HE}) becomes 
\begin{equation}\label{eq:sigma:sur:p:parametrique:HE:continuous}
\frac{\sigma_p}{p} = 
\frac{4}{N^{1/6} \sqrt{2L}} 
\left( \frac{p}{p_0} \right)^{1/3}
\left( \sigma^2 X_0 \right)^{1/6},
\end{equation}
that we can compare to the cell-optimization expression (eq. (12) of
\cite{Bernard:2012uf}):

\begin{equation}
\frac{\sigma_p}{p} = 
\frac{C}{\sqrt{2L}} 
\left( \frac{p}{p_0} \right)^{1/3}
\left( \sigma^2 X_0 \right)^{1/6}
\label{eq:resolution:min}
\end{equation}
with $C \equiv 5^{1/6} + 5^{-5/6} \approx 1.57 $.
We see that the precisions are commensurate at small $N$ and that the
present approach becomes more precise at larger $N$, within the
high-momentum approximation, $p \gg p_s$.
 
\subsection{Cramér-Rao Bounds}
\label{sub:sec:Cramer:Rao:bound}

The Cramér-Rao bound is a lower bound on the variance of an estimator.
If the variance of the estimator reaches the Cramér-Rao bound, it can
be stated that the estimate is optimal.
The Cramér-Rao criterion for an estimator $\hat \theta$ of a parameter
$\theta$ obtained from measurements $Z^N$ is \cite{Matisko_Cr:2012}:
\begin{equation}\label{eq:I}
I(\theta)=-\E\left(\partial_\theta\left[\partial_\theta p(Z^N|\theta)\right]\right),
\end{equation}
where $I$ is the Fischer information.
If $\hat \theta$ is an unbiased estimator of $\theta$, then
\begin{equation}\label{eq:Cramer:Rao}
\E\left((\hat \theta-\theta)^2\right)\ge I^{-1}(\theta).
\end{equation}

Following the recursive method of \cite{Matisko_Cr:2012} we obtain 
\begin{equation}\label{eq:I:final}
I(s)=\gfrac{N}{2 s^2} ,
\end{equation}
that is, finally, the obvious 
\begin{equation}\label{eq:CR:precision}
\gfrac{\sigma_p}{p} \ge \gfrac{1}{\sqrt{2 N}}.
\end{equation}

No major insight obtained with the Cramér-Rao Bounds then.

\subsection{Smoothing and Momentum Measurement}
\label{sub:sec:smoothing}

In this section \ref{sec:momentum:measurement} we have obtained an
optimal estimator of a charged particle momentum based on the
analysis of the filtering innovations of KFs with variable $s$
parameters.
After filtering, a KF provides an optimal estimate of the state vector parameters
(transverse position and angle) of the track at the end of the track.
An optimal estimate all along the track can be obtained by an additional,
backward, pass named smoothing \cite{Fruhwirth:1987fm}.
One might consider a scheme for momentum measurement based on the
smoothing innovations rather than on the filtering innovations in the
hope that the performance would be even better.

Actually smoothing is equivalent to an optimal linear combination of
two independent filterings performed in the direct and in the backward
directions, respectively \cite{Fraser}.
We have compared the values of the estimators
of the particle momentum obtained in the two directions and found them
to be equal for each track.
Therefore no further improvement is to be expected with such a combination,
nor with a measurement based on smoothing innovations.

\section{Conclusion}
\label{sec:conclusion}

We first reconsider tracking with multiple scattering and detector
resolution in magnetic-field-free detectors under the assumption that
the track momentum is known, using optimal methods.
This is done under a number of approximations, including
Gaussian-distributed multiple-scattering deflections and the absence
of energy loss during propagation.
The information matrix is updated recursively while the track 
proceeds through the detector: after this mechanism has converged,
the information matrix is found to be a solution of a Riccati equation which is not
surprising as this optimal estimation can be performed with a Kalman
filter.

For segmented detectors (discrete Riccati equation) and homogeneous
detectors (continuous Riccati equation), we obtain exact expressions
of the variances of the intercept and of the angle from the solution
of that equation (eqs. (\ref{eq:exact:discret}) and
(\ref{eq:CARE:exact}), respectively).
We compare their Taylor expansions with expressions published in the past.
Convergence ({\bf thick detector}) takes place after a detector thickness
$L \gtrsim 2.5 \lambda$ for homogeneous detectors
($l \lesssim 0.2 \lambda$), and for somewhat larger values of $L$ for
segmented detectors (Fig. \ref{fig:liminf}).
$\lambda$ is the detector scattering length for track
 momentum $p$.

For a given track momentum, a {\bf homogeneous detector} is defined as the
small longitudinal sampling limit, $l \to 0$,
$\imath$ and $s$ being kept constant.
In practice a limit of $l/\lambda \lesssim 0.2$ is found (Fig. \ref{fig:V_x}).
In contrast with magnetic spectrometers, for which the large $L/\lambda$
Taylor expansion contains $1/(L/\lambda)^n$ terms, here ($\vec B = \vec 0$), the
expansion contains only exponential terms and convergence is therefore
much faster
(Fig. \ref{g:WrapUp}).
For coarse segmented detectors for which $l/\lambda \gtrsim 2$,
e.g. for $p \lesssim 35\,\mega\electronvolt/c$ for eASTROGAM
\cite{E-Astrogam:2016} or AMEGO \cite{AMEGO} a KF becomes useless as
the angular resolution is determined mainly by the measurements in the
two first wafers (Fig. \ref{g:lambda}).

We then obtain an optimal estimator of the track momentum by a
Bayesian analysis of the filtering innovations of a series of Kalman
filters applied to the track.
A numerical characterisation of the method shows that for a given
detector the method is reliable up to some limit momentum $p_\ell$
above which the relative precision $\sigma_p / p$ becomes larger than unity.
For lower momentum tracks, $p \ll p_\ell$, the momentum estimation is
found to be unbiased.
We perform a parametric study of the estimator from which we extract a
heuristic analytical description of the relative uncertainty of the
momentum measurement (eq. (\ref{eq:sigma:sur:p:parametrique})).

\section{Acknowledgment}

It is a pleasure to acknowledge the support of the French National
Research Agency (ANR-13-BS05-0002).

\clearpage

\begin{center}
 \footnotesize
\begin{tabular}{llll}
$\alpha$ & one eigenvalue of $\Phi$ with norm larger than 1 & eqs. (\ref{eq:Phi:eigenspectrum}), (\ref{eq:alpha:of:x}) \\ 
$a$ & charged particle ``intercept'' at vertex & eq.~(\ref{eq:droite}) \\ 
$b$ & charged particle slope at vertex & eq.~(\ref{eq:droite}) \\ 
$A_n$ & matrix used in the calculation of $I_n$ & Subsec. \ref{sub:sec:segmented:detector}, see eq.~(\ref{eq:fonctio}) \\ 
$B_n$ & matrix used in the calculation of $I_n$ & Subsec. \ref{sub:sec:segmented:detector}, see eq.~(\ref{eq:fonctio}) \\ 
$B$ & scattering matrix & eq.~(\ref{eq:Billoir:scattering}) \\ 
$B$ & magnetic field & Sec. \ref{sec:introduction} \\ 
$\beta$ & charged particle velocity normalized to the velocity of light in vacuum & Sec. \ref{sec:introduction}, see eq.~(\ref{eq:multiple:scattering:base}) \\ 
$\beta$ & scale factor & Subsec. \ref{sub:sec:segmented:detector}, see eq.~(\ref{eq:AB:Norm:beta}) \\ 
 $\beta_n$ & Kalman innovation probability density & Subsec. \ref{sub:sec:bayesian:method}, see eq.~(\ref{eq:pn:3}) \\
 $\delta$ & neutrinos: CP-violating complex phase of the PMNS matrix & Sec. \ref{sec:introduction} \\
 $D$ & drift matrix from layer $n$ to layer $n+1$ & eq.~(\ref{eq:Billoir:drift}) \\
$\Delta x$ & active target material thickness through which multiple scattering takes place & Sec. \ref{sec:introduction}, see eq.~(\ref{eq:multiple:scattering:base}) \\ 
$E$ & photon energy & Sec. \ref{sec:introduction} \\
$\E$ & expectation value & Sec. \ref{sec:kalman} \\
$\gamma$ & charged particle Lorentz factor & Sec. \ref{sec:introduction} \\ 
$H$ & Kalman measurement matrix & Sec. \ref{sec:kalman}, see eq.~(\ref{eq:zn}) \\
$j$ & the imaginary unit & Subsec. \ref{sub:sec:segmented:detector} \\
$\imath$ & measurement information density per unit track length & Sec. \ref{sec:tracking}, see eq.~(\ref{eq:Billoir:measurement}) \\
$I$ & information matrix & eq.~(\ref{eq:Billoir}) and eq.~(\ref{eq:I}) \\
$J$ & a constant matrix & eq.~(\ref{eq:def:J} \\
$K$ & Kalman gain matrix & eq.~(\ref{eq:Kn}) \\
$L$ & total detector thickness & Sec. \ref{sec:tracking} \\ 
$l$ & space between two successive detector layers & Sec. \ref{sec:tracking}, see eq.~(\ref{eq:Billoir:drift}) \\ 
$\lambda$ & detector scattering length at momentum $p$ & eq.~(\ref{eq:def:lambda}) \\ 
$M$ & measurement matrix & eq.~(\ref{eq:Billoir:measurement}) \\
$\nu_n$ & Kalman innovations & eq.~(\ref{eq:nun}) \\
$n$ & layer index & Sec. \ref{sec:tracking} \\ 
$N$ & number of layers in detector & Sec. \ref{sec:tracking} \\
$\mathcal N$ & normal or Gaussian probability density & eq.~(\ref{eq:p:normal}) \\
$\Phi$ & matrix that performs the transformation from
$\begin{bmatrix} A_{n} \\ B_{n} \end{bmatrix}$
to
$\begin{bmatrix} A_{n+1} \\ B_{n+1} \end{bmatrix}$ & eq.~(\ref{eq:def:Phi}) \\ 
 $p$ & probability density & Sec. \ref{sec:kalman}, see eq.~(\ref{eq:pn})\\
$p$ & charged particle momentum & Sec. \ref{sec:introduction} \\ 
$p_0$ & multiple scattering constant & Sec. \ref{sec:introduction}, see eq.~(\ref{eq:multiple:scattering:base}) \\
\end{tabular}
\end{center}

\begin{center} 
 \footnotesize
\begin{tabular}{llll}
$p_1$ & detector tracking angle resolution characteristic momentum & eq.~(\ref{eq:p_1}) \\
$p_u$ & detector thin/thick limit momentum & eq.~(\ref{eq:pu}) \\
$p_x$ & detector homogeneous/segmented limit momentum & eq.~(\ref{eq:px}) \\
$p_s$ & detector limit momentum between the $\sigma_p/p = 1/\sqrt{2N}$ and $\sigma_p/p \propto p^{1/3}$ ranges & eq.~(\ref{eq:ps}) \\
$p_\ell$ & detector limit momentum for which $\sigma_p/p = 1$ & eq.~(\ref{eq:pell}) \\
$P_n$ & Kalman state covariance matrix & Sec. \ref{sec:kalman}, see eq.~(\ref{eq:Pn}) \\
$q$ & particle electric charge & Sec. \ref{sec:introduction} \\ 
$\rho$ & charged particle trajectory curvature radius & Sec. \ref{sec:introduction} \\ 
$\sigma$ & single-track single-layer space resolution & Sec. \ref{sec:introduction} \\ 
$\sigma_p$ & momentum resolution & Sec. \ref{sec:introduction} \\ 
$s$ & average multiple scattering angle variance per unit track length & Sec. \ref{sec:tracking}, see eq.~(\ref{eq:Billoir:scattering}) \\ 
$S_n$ & Kalman innovation covariance matrix & eq.~(\ref{eq:Sn}) \\
$\theta$ & a parameter & Sec. \ref{sub:sec:Cramer:Rao:bound}, see eq.~(\ref{eq:I}) \\
$\hat \theta$ & estimator for parameter $\theta$ & Sec. \ref{sub:sec:Cramer:Rao:bound}, see eq.~(\ref{eq:I}) \\
$\theta_0$ & multiple scattering RMS angle & eq.~(\ref{eq:multiple:scattering:base}) \\ 
$u$ & detector thickness normalized to detector scattering length at momentum $p$ & eq.~(\ref{eq:def:u}) \\
$u_n$ & deflection angle & Sec. \ref{sec:kalman}, see eq.~(\ref{eq:propagation:xn}) \\
$v_n$ & Kalman measurement noise & Sec. \ref{sec:kalman}, see eq.~(\ref{eq:zn}) \\
$V$ & particle state vector (``intercept'', angle) correlation matrix & Sec. \ref{sec:tracking}, see eq.~(\ref{eq:Billoir}) \\ 
$X_0$ & active target material radiation length & Sec. \ref{sec:introduction}, see eq.~(\ref{eq:multiple:scattering:base}) \\ 
$x$ & detector longitudinal sampling normalized to scattering length at momentum $p$ & Subsec. \ref{sub:sec:segmented:detector}, see eq.~(\ref{eq:alpha:of:x}) \\ 
$x$ & axis name \\
$x_n$ & Kalman state vector & eq.~(\ref{eq:propagation:xn}) \\
$X$ & matrix used in the calculation of $\Phi'$ & eq.~(\ref{eq:contexp}) \\ 
$Y$ & matrix used in the calculation of $\Phi'$ & eq.~(\ref{eq:contexp}) \\ 
$y$ & axis name \\
$z_n$ & Kalman measurements & eq.~(\ref{eq:zn})\\
$z$ & axis name \\
$Z$ & active target atomic number & Sec. \ref{sec:introduction} \\ 
$Z_n$ & set of measurements, $z_0 \cdots z_n$ & Sec. \ref{sec:kalman}, see eq.~(\ref{eq:pn})\\
\end{tabular}
\end{center}

\clearpage

\tableofcontents

\end{document}